\documentclass[%
 prc,
superscriptaddress,
 amsmath,amssymb,
 aps,twocolumn
]{revtex4-2}
\usepackage{epsfig} 
\usepackage{bmpsize}
\usepackage{mathtools}
\usepackage{color}
\usepackage{array}   
\newcolumntype{L}{>{$}l<{$}} 
\newcommand{\bracket}[2]{\left< #1 | #2 \right>}
\newcommand{\bra}[1]{\left< #1 \right|}
\newcommand{\ket}[1]{\left| #1 \right>}
\newcommand{\mv}[1]{\left< #1 \right>}
\newcommand{\mf}[1]{\mathbf{#1}}
\newcommand{\mfr}{\mathbf{r}}
\newcommand{\kap}{\kappa}
\newcommand{\+}{\dagger}
\newcommand{\up}{\uparrow}
\newcommand{\dn}{\downarrow}
\newcommand{\dd}[1]{\partial_{#1}}
\newcommand{\pd}[2]{\partial_{#2} #1}
\newcommand{\dy}{~$\prescript{132}{66}{\textrm{Dy}}$~}
\newcommand{\gd}{~$\prescript{132}{64}{\textrm{Gd}}$~}
\newcommand{\nd}{~$\prescript{132}{60}{\textrm{Nd}}$~}

\begin{document} 
 
 \title{Symmetry properties of pair correlations in deformed heavy nuclei}

\author{Georgios Palkanoglou}
\affiliation{TRIUMF, 4004 Wesbrook Mall, Vancouver, BC V6T 2A3, Canada}
\affiliation{Department of Physics, University of Guelph, 
Guelph, ON N1G 2W1, Canada}
\author{Alexandros Gezerlis}
\affiliation{Department of Physics, University of Guelph, 
Guelph, ON N1G 2W1, Canada}

\begin{abstract} 
Nucleons are known to form pairing correlations with various types of spin-symmetries. Spin-singlet neutron-neutron and proton-proton pairing is abundant in the nuclear chart but spin-triplet and mixed-spin proton-neutron pairing correlations have also been predicted to form at least in the ground states of certain nuclei. A realistic candidate region is that of the lightest Lanthanides where it was recently demonstrated that the nuclear deformation expected to emerge enhances spin-triplet pairing correlations. In this paper we provide the details of the deformed multimodal Hartree-Fock-Bogolyubov theory that lead to this conclusion, as well as the details of the effects identified. We present in detail the response of different pairing correlations to various deformation modes and calculate their signatures in the odd-even staggering of masses. This paper provides a detailed discussion, and some resolutions, on the long-standing question ``what is the effect of nuclear deformation on the various pairing correlations?''
\end{abstract} 

\maketitle 
\section{Introduction}
Pairing correlations are a ubiquitous property of fermionic systems and the underlying mechanism of fermionic superfluidity. Nucleons, owing to the rich structure of the nuclear force, have been known to display novel types of pairing correlations, both in extended as well as finite systems~\cite{Dean:2003}. This coupled with the inherent difficulty in describing strongly interacting fermions and our partial understanding of the nuclear force, has kept nuclear superfluidity an active field of research~\cite{Soma:2011,Palkanoglou:2022,Barbieri:2021,Drissi:2024,Schuck:2019,Gupta:2024,Martin:2016,Sourie:2021,Rios:2017,Sedrakian:2019,Niu:2018, Magierski:2017, Bernard:2019, Oishi:2017, Scamps:2012, Sadhukhan:2014}.

In the case of extended matter, where the pairs formed condense and one can unambiguously talk about nuclear superfluids, the standard example is the crust of neutron stars where neutrons have been known to exhibit $^1S_0$ superfluidity but also the more exotic $^3P_2-^3F_2$ superfluidity driven mainly by the tensor part of the neutron-neutron interaction. In proton-neutron mixtures, commonly referred to as nuclear matter and expected to arise in the neutron-star interior, the existence of deuteron-type superfluidity, with pairing between protons and neutrons, has also been posited~\cite{bertsch:2011}. 

In the case of finite systems, typically atomic nuclei, the finite-size effects complicate the description. From the outset the fundamental problem of defining a macroscopic condensate with a small number of particles introduces an ambiguity that forces one to largely give up the language of condensates and focus on the correlations present in the many-body wavefunction and their associated phenomenology. In this language, spin-singlet pairing correlations between like particles are seen in the increased binding of nuclei with even numbers of protons and neutrons and in two-particle transfer reactions' cross sections; this was historically the first type of pairing correlations identified, predating many advancements in the field of nuclear pairing physics. Like-particle spin-triplet pairing correlations have also been proposed to form in medium-to-heavy nuclei, driven by the attraction of the nuclear interaction in the $^3\textrm{P}_2$ channel, reminiscent of the $^3\textrm{P}_2-^3\textrm{F}_2$ condensate in neutron matter~\cite{hinohara:2024}. Spin-triplet proton-neutron pairing correlations are expected to dominate in nuclei with equal numbers of protons and neutrons large enough to not be dominated by surface effects, with radii larger than $\sim9~\textrm{fm}$, or with nuclear structure allowing the paired particles to move away from the nuclear surface~\cite{poves:1998,baroni:2010,bertsch:2011}. For these types of correlations various phenomena have been proposed with varying degrees of identifiability: suppression of the odd-even staggering effect~\cite{bertsch:2011}, the nucleus' response to rotation, deuteron-transfer reactions' cross sections, etc~\cite{Frauendorf:2001, Isacker:2021,Matsubara:2015,frauendorf:2014}. Because this type of pairing has to be found in heavy nuclei, where other nuclear structure phenomena can be overwhelming, none of these signatures has been seen experimentally yet. Finally, mixed-spin proton-neutron pairing correlations might form in large nuclei with small mismatches in proton and neutron numbers, that is, in large finite nuclear systems where neither spin-singlet nor spin-triplet pairing correlations dominate~\cite{gezerlis:2011,bulthuis:2016,rrapaj:2019}. Since these types of correlations have to be found in systems where spin-triplet pairing is non-negligible, any experimental exploration is plagued by similar issues. Given their exotic character that limits the comparisons with other pairing correlations, the phenomenology of mixed-spin proton-neutron pairing is still underdeveloped.

At present, most of the understanding of the phenomenology of pairing correlations in nuclei, especially those of the spin-triplet and mixed-spin type, comes from mean-field models~\cite{frauendorf:2014} as these nuclei have historically posed a challenge for first-principle, i.e., \textit{ab initio}, approaches. This is because these types of correlations are expected to be most dominant in heavy nuclei whose \textit{ab initio} description has been challenging~\cite{Scalesi:2024,Tichai:2018,Hergert:2018}. Additionally, the relevant nuclei display moderate deformation that is in itself challenging to describe from first principles and proximity to the proton-drip line in the nuclear chart that necessitates the inclusion of the continuum degrees of freedom in the description. The latter is currently an onerous task even for the lightest nuclides~\cite{baroni:2013}. The mean-field descriptions available for these systems often ignore their deformed character and in those that can describe deformation, the description is either limited to certain deformation modes, e.g., only quadrupole~\cite{gambacurta:2015}, or only allowing for the possibility of spin-singlet pairing~\cite{marevic:2022,doba:2021}. In other words, a complete description of spin-singlet, spin-triplet, and mixed-spin pairing correlations in the relevant deformed heavy nuclei has been missing. 

In this paper, we provide the details of a self-consistent mean-field description of deformed heavy nuclei including pairing effects in spin-singlet, spin-triplet, and mixed-spin channels. We introduced this description in Ref.~\cite{Palkanoglou:2025} and used it to conclude that the existence of spin-triplet and mixed-spin pairing correlations in the region of light Lanthanides of the nuclear chart was not an artifact of neglecting the moderate deformation that is expected for these nuclei. In this paper we provide a detailed description of the formalism and an in-depth analysis of the pairing's response to deformation. The rest of this paper is organized as follows: in Sec. \ref{sec:deform} we provide the details on the way the deformation is modeled, in Sec. \ref{sec:axialhfb} we describe the mean-field description based on a Hartree-Fock-Bogolyubov (HFB) approach, and in Sec. \ref{sec:study} we calculate correlation energies associated with different pairing modes and odd-even staggerings, and we study in detail the pairing correlations formed in the ground states of deformed nuclei along with any associated phenomena.

\section{Modeling axial deformation}
\label{sec:deform}
A mean-field description of nucleons in a nucleus amounts to assuming that each particle moves in a static configuration composed of the rest of the nucleus. In this phenomenological picture, the shape of the nucleus is put in by-hand as the surface of the one-body potential felt by each particle. When describing spherical nuclei, this is achieved trivially, in the case of non-spherical shapes, one needs to parametrize the nuclear surface in an efficient way. In any case, one starts from defining the surface of the nucleus as the points $\mf{r}$ satisfying an equation of the form
\begin{align}
    \Phi(\mf{r})=0~,
\end{align}
for some scalar function $\Phi$ which we define using Cassini ovals.
Cassini ovals were first developed in the context of planetary astronomy, but they have been used extensively in nuclear physics to parametrize the nuclear surface~\cite{Garrote:1995,Stravinsky:1968,Moller:1972,Okada:2023,Carjan:2012} and they have been especially successful in describing the extreme deformations appearing in fissioning and scissioning nuclei. In what follows we provide an overview of modeling nuclear deformation with Cassini ovals; more details can be found in \cite{Stravinsky:1968}.

The main idea behind the use of Cassini ovals to describe the nuclear surface is that one can choose a system of coordinates whose coordinate surfaces (i.e., the surfaces generated by fixing one coordinate and varying the rest) are good approximations of the surfaces of deformed nuclei. In the presence of only axial deformation, the nuclear shape is symmetric under rotations around one axis and so it's sufficient to describe the surface on the $(\rho,z)$ plane of the cylindrical coordinates:
\begin{align}
\Phi(\rho,z) &= R-R(x) \\ 
&=\left[(\bar{z}^2+\bar{\rho}^2)^2-2\varepsilon R_0^2(\bar{z}^2+\bar{\rho}^2)+\varepsilon^2R_0^4\right]^{1/4}- \notag \\ 
&- R_0\left[1+\sum_{\lambda\neq0}^{\lambda_{\textrm{max}}} \beta_\lambda P_\lambda(x)\right]~.
\end{align} 
The barred coordinates are shifted cylindrical coordinates to account for the incompressibility of a nucleus with mass number $A$,
\begin{align}
\bar{\rho}=c\rho~, \quad \bar{z}=cz+\bar{z}_{\rm cm}~, \quad R_0=\rho_0 A^{1/3} ~,
\end{align} 
and $R$ and $x$ are the coordinates in the plane of the Cassini ovals:
\begin{align}
R &= \left[(\bar{z}^2+\bar{\rho}^2)^2-2\varepsilon R_0^2(\bar{z}^2+\bar{\rho}^2)+\varepsilon^2R_0^4\right]^{1/4} \\
x &= \frac{\textrm{sgn}(\bar{z})}{\sqrt{2}} \left\{1+\frac{\bar{z}^2-\bar{\rho}^2-\varepsilon R_0^2}{\left[(\bar{z}^2+\bar{\rho}^2)^2-2s(\bar{z}^2-\bar{\rho}^2)+\varepsilon^2R_0^4\right]^{1/2}}\right\}!. \label{eq:cassini}
\end{align}
In other words, we describe the surface of the nucleus as a Cassini oval, characterized by the elongation $\varepsilon$ with additional multipole deformation, characterized by the parameters $\{\beta_\lambda\}$, 
\begin{align}
R(x)=R_0[1+\sum_{\lambda\neq 0}^{\lambda_{\textrm{max}}}\beta_\lambda P_\lambda(x)]   ~, \label{eq:std_expan}
\end{align}
using the Legendre polynomials $P_\lambda$. In the absence of elongation ($\varepsilon=0$), the $x=\cos\theta=\textrm{const.}$ lines are straight lines radiating from the origin, and the corresponding Cassini oval is a circle. In that case the deformation of the nuclear surface is any multipole deformation perscribed by the parameters $\beta_\lambda$.

Having defined the nuclear surface via the function $\Phi(\rho,z)$, the distance of any point $(\rho,z)$ from that surface can be approximated by
\begin{align}
L(\rho,z) = \frac{\Phi(\rho,z)}{\left|\nabla \Phi(\rho,z)\right|}~. \label{eq:dist}
\end{align}
This definition is based on a few standard assumptions. First, the nucleus' surface diffuseness is assumed to be constant along the surface and independent of deformation. This is supported by the short-range of the nuclear force and it is equivalent to assuming that the potential's derivative normal to the equipotential curves has a constant magnitude along the nuclear surface. Additionally, the potential is assumed to reach a constant non-zero value at the center of the nucleus and vanish at infinity. Both of these assumptions are consequences of the saturation of nuclear forces which allows us to describe the mean-field potential as an image of the nuclear density profile in the first place. Expanding the distance function in Eq.~(\ref{eq:dist}) in the normal displacement to the surface $l$, one gets:
\begin{align}
L = l + \gamma l^2 + O(l^4)~,
\end{align}
where $\gamma$ depends on the specific shape of the nucleus and it's of the order of the average curvature of the surface. This term remains negligible for small diffuseness, which is again prescribed by the short-ranged nucleon-nucleon interaction. In this case $L$ can be seen as shape-independent and used to measure distances from the surface and to define the average nuclear potential as a generalization of the standard Wood-Saxon form,
\begin{align}
V_{\textrm{WS}}(\rho,z) = V_0 \left[1+e^{L(\rho,z)/a}\right]^{-1}~, \label{eq:pot_ws}
\end{align}
where $a$ stands for the aforementioned diffuseness of the nuclear surface and $V_0$ is a constant taken from Ref.~\cite{book:bohrmottelson}. The choice of the distance function $L$ turns this potential to the standard Wood-Saxon potential in the absence of deformation.

While modeling nuclear deformation with Cassini ovals is most beneficial when looking at elongated nuclei, for the results of this paper we keep $\varepsilon=0$ because that allows us to connect with the rest of the literature in a well-defined way. To see why, we note that for given elongation $\varepsilon$, the $x$ coordinate of the Cassini plane can be seen as a function of $\bar{r}=\sqrt{\bar{\rho}^2+\bar{z}^2}$ and $\cos \theta$, with the $\bar{r}$ dependence dropping out at $\varepsilon=0$. Hence, the Legendre polynomial in Eq.~(\ref{eq:std_expan}) can be written as
\begin{align}
P_{\lambda}\left[x(\bar{r},\cos\theta)\right] = \sum_{\alpha\beta} y_{\alpha\beta}^{\lambda}(\bar{r})Y_{\alpha\beta}(\cos \theta)
\end{align}
With these, the expression for the radius becomes
\begin{align}
R\left[x(\bar{\rho},\bar{z})\right]&=R_0\sum_{\lambda\mu} \alpha_{\lambda}P_{\lambda}(x) \\
&= R_0\sum_{\substack{\lambda\mu \\\alpha\beta}} \alpha _{\lambda\mu}y_{\alpha\beta}^{\lambda\mu}(\bar{r})Y_{\alpha\beta}(\cos \theta)~,
\end{align}
where
\begin{align}
y_{\alpha\beta}^{\lambda\mu}(\bar{r}) &= \int d(\cos \theta) P_{\lambda}\left[x(\bar{r},\cos\theta)\right]Y^*_{\alpha\beta}(\cos \theta)~.
\end{align}
Comparing this to the usual expansion of an axially symmetric nuclear surface to multipoles, $R(\theta) = \sum_{\lambda} \beta_\lambda P_{\lambda}(\cos \theta)$, we see that $\varepsilon\neq 0$ generates a $\bar{r}$-dependence in the expansion coefficients $\{\beta_\lambda\}$. That is, $\varepsilon$ generates deformation beyond that of multipoles. In Eq.~(\ref{eq:pot_ws}), one could absorb this dependence in additional $\rho$-dependence departing from the common Woods-Saxon form. However, that could be problematic when comparing with deformation parameters found from calculations and experiments as those are usually connected to the deformation parameters in Eq.~(\ref{eq:std_expan}) in a Wood-Saxon form. Hence, with the interest of keeping with the standard Wood-Saxon form and not large deformations, we set $\varepsilon=0$ for the rest of this paper. Here by large deformation we mean the shapes appearing in, e.g., scissioning nuclei which can be described by Cassini ovals with $\varepsilon\sim 1$; these are topics left for future work. 

\section{Axially symmetric HFB with multiple pairing symmetries}
\label{sec:axialhfb}
In this section we describe a formulation of the standard HFB approach to describe deformed nuclei with any type of axial symmetry and various types of pairing spin symmetry -- spin-singlet, spin-triplet, or mixed-spin. 

Before focusing on  the individual ingredients of our formulation, we will go over the standard HFB approach. There a Hamiltonian containing one- and two-body terms can be expressed in second quantization:
\begin{equation}
H=\sum_{ij}\epsilon_{ij}c_i^\+ c_{j} + \sum_{i>j, k>l} \bra{ij}v\ket{kl} c_i^\+  c_j^\+ c_l c_k~, \label{eq:ham}
\end{equation}
using the creation (and annihilation) operators $c_i^\+$ ($c_i$) for an assumed single-particle basis $\ket{i}=c_i^\+\ket{0}$ to be specified later. From this basis one then constructs the Bogolyubov quasi-particle basis,
\begin{align}
    \alpha^\+_j=\sum_{i} U_{ij}c_i^\+ + V_{ij}c_i~\label{eq:quasibasis}
\end{align}
defined via the matrices $U$ and $V$ and the quasi-particle vacuum, i.e., the state with no quasi-particles:
\begin{align}
    \ket{\textrm{HFB}} \propto \prod_i \alpha_i \ket{0}~,\label{eq:hfbstate}
\end{align}
where $\ket{0}$ is the particle vacuum. As evident by the notation, this state is commonly called an \textit{HFB state} and it can be uniquely defined, for a set of $U$ and $V$ matrices, because the basis $\{\beta_i\ket{0}\}$ is an orthonormal basis (assuming that the starting basis $\{c_i\ket{0}\}$ is also orthonormal). The expectation value of the Hamiltonian on any HFB state will depend on the matrices $U$ and $V$, and the choice that yields the lowest energy defines the HFB ground state. A defining feature of HFB states is a breaking of certain of the Hamiltonian's symmetries: owing to mixing of creation and annihilation operators in each element of the quasiparticle basis in Eq.~(\ref{eq:quasibasis}), some symmetries of the original basis $\{c_i\ket{0}\}$ are lost. These are two $U(1)$ symmetries associated with the particle number (one for neutrons one for protons), and one $SU(2)$ associated with the angular momentum. In other words, each HFB ground state is a linear combination of states, each with good particle and angular momentum quantum numbers, yielding a distribution of particle numbers and angular momenta. These broken symmetries can then be conserved \textit{on average} by centering these distributions around the desired quantum numbers. This is done by adding the corresponding operators to Eq.~(\ref{eq:ham}) multiplied by Langrange multipliers, which are then adjusted appropriately.

In practice, the HFB ground state, that is, the matrices $U$ and $V$, can be identified using the method of gradient descent. In this one starts with an assumption for the HFB ground state, that is, a state $\ket{\Phi}$ in the form of an HFB state in Eq.~(\ref{eq:hfbstate}) with some matrices $U$ and $V$, and performs a ``walk'' in the space of states non-orthogonal to $\ket{\Phi}$ using Thouless' theorem which states that any state non-orthogonal to $\ket{\Phi}$ can be written as
\begin{align}
    \ket{\Phi'}= e^{\sum_{i<j}Z_{ij}\alpha_i^\+\alpha_j^\+}\ket{\Phi}~.
\end{align}
If one then ensures that each step lowers the energy, i.e., performing a gradient descent, the walk will converge to the HFB ground state. One major advantage of this method of finding the HFB ground state is that we can constrain the ``walk'' by ``penalizing'' states with certain properties. This can be done by adding the appropriate terms in the Hamitlonian multiplied by Lagrange multipliers which can then be adjusted to target states where the associated operators have the desired values. We use this to target specific average proton and neutron numbers when looking at the ground state of a nucleus, as well as, specific amplitudes of the pairing fields when looking at HFB states with specific pairing symmetries. More details on the latter are given below. 

We will now describe the specific adjustments to the standard HFB approach that allow us to tackle axially deformed nuclei with spin-triplet, spin-singlet, and mixed-spin pairing. In order, we will describe: (A) how we use the Cassini ovals introduced in sec. \ref{sec:deform} to provide the appropriately deformed mean-field and the associated basis $\{c_i\}$ to be used in Eq.~(\ref{eq:quasibasis}), (B) how we define the two-body term in Eq.~(\ref{eq:ham}), and what that means for the two-body quantities of HFB theory, to describe pairing between six possible types of pairing defined by the spin and isospin quantum numbers of the pair, and (C) how we define proxies for the symmetries of the pairing correlations in an HFB state.

\subsection{Axially symmetric mean-field Hamiltonian}
The one-body part, $\epsilon$, contains the single-particle kinetic energy, a Wood-Saxon potential deformed appropriately to match the axially-symmetric density profile of the nucleus, and a spin-orbit term ($B$) with a space-dependent factor that peaks at the deformed nuclear surface: 
\begin{align}
\epsilon&=h + B~, \\
h&= \frac{p^2}{2m} + V_{WS} (\rho,z)~, \\
B&=\frac{1}{2}\kappa\boldsymbol{\nabla}f(\mfr)\cdot \left(\boldsymbol{\sigma}\times \mf{p}\right)~,
\end{align}
where $\kappa$ is a constant and $f(\mfr)$ the function in the square brackets in Eq.~(\ref{eq:pot_ws}). We represent $\epsilon$ on the basis $\ket{i}$ defined by the eigenvectors of $h$ in cylindrical coordinates:
\begin{align}
\epsilon_{ij} &= \bra{i}\epsilon\ket{j} = h_i \delta_{ij} + B_{ij}~, \label{eq:hfb_onebody} \\
B_{ij}&=\delta_{l_{zi}l_{zj}}  I^{(l_{zi})}_{n_i n_j s_{zi} s_{zj}}~, \label{eq:Bso}
\end{align}
where $n_i$ and $l_{zi}$ are the radial and angular quantum numbers, respectively, of state $i$.
\begin{align}
I^{(j_{z_i})}_{n_in_j s_{z_i} s_{z_j}}&= \int d \rho dz \,u^*_{n_i l_{zi}}(\rho,z)   B_{s_{zi}s_{zj}}(\mf{r}) u_{n_jl_{zj}} (\rho,z) \\
I^{(j_{zi})}_{n_in_j\up\up} &= -\frac{\kap}{2} \int d\rho dz u^*_{n_i l_{zi}}(\rho,z) \pd{f}{\rho} \frac{j_{zi}}{\rho} u_{n_jl_{zi}}(\rho,z)~, \label{eq:i11} \\
I^{(j_{zi})}_{n_in_j\up\dn} &= -\frac{\kap}{2}\int d\rho dz u^*_{n_il_{zi}}(\rho,z)\bigg[\pd{f}{\rho}\dd{z} - \pd{f}{z}\dd{\rho} - \notag\\
&\quad\quad - \pd{f}{z}\frac{j_{zi}}{\rho}\bigg]u_{n_jl_{zi}}(\rho,z)~,  \label{eq:i12}\\
I^{(j_{zi})}_{n_in_j\dn\up} &= \frac{\kap}{2}\int d\rho dz u^*_{n_i l_{zi}}(\rho,z)\bigg[\pd{f}{\rho}\dd{z} - \pd{f}{z}\dd{\rho}+ \notag\\
&\quad\quad+ \pd{f}{z}\frac{j_{zi}}{\rho} \bigg]u_{n_j l_{zi}}(\rho,z)~,  \label{eq:i21}\\
I^{(j_{zi})}_{n_in_j\dn\dn} &=  \frac{\kap}{2} \int d\rho dz u^*_{n_i l_{zi}}(\rho,z)\pd{f}{\rho} \frac{j_{zi}}{\rho} u_{n_jl_{zi}}(\rho,z)~, \label{eq:i22}
\end{align}
where we use indices of $i$ and $j$ to signify single-particle states whose wavefunctions $\psi_i(\rho,\varphi,z)=(1/2\pi)u_{n_i l_{zi}}(\rho,z)e^{il_z\varphi}/\sqrt{\rho}\chi(s_{zi})$ are found by solving the one-body Schr{\"o}ndiger's equation for $h$ in cylindrical coordinates~\cite{Gordon:2006}. 

\subsection{The pairing interaction}
\label{sec:pairinter}
The second term in Eq.~(\ref{eq:ham}) is the two-body term commonly identified with the pairing interaction, that is, the interaction that is responsible for the creation of the pairing correlations. We consider a zero-range pairing interaction,
\begin{align}
v(\mf{r},\mf{r}') = \sum_\alpha v_\alpha \delta^{(3)}(\mf{r}-\mf{r}')P_{J_z=0}P_{\alpha}~, \label{eq:pairint}
\end{align}
where the $v_\alpha$ determine the strength of the pairing interaction in 6 distinct spin-isospin channels labeled by $\alpha$ and shown in Table~\ref{tb:channels}
\begin{table}[]
    \centering
    \begin{tabular}{|L|LL|LL|}
\hline
\alpha & S&(S_z) & T&(T_z)\\
\hline
\hline
0 & 0& & 1&(1) \\
1 & 0& & 1&(0) \\
2 & 0& & 1&(-1) \\
3 & 1&(1) & 0 & \\
4 & 1&(0) & 0 & \\
5 & 1&(-1) & 0 & \\
\hline
\end{tabular}
    \caption{The spin-isospin channels of the pairing correlations that we consider.}
    \label{tb:channels}
\end{table}
and implemented via the projection operator $P_\alpha$. That is, we consider 6 types of pairing correlations: 3 with spin-singlet symmetry (also called isovector) and 3 with spin-triplet symmetry (also called isoscalar), i.e, pair wavefunctiosn that behave like scalars or vectors, respectively. under spin-rotations. More importantly, we only account for spin-triplet pairing correlations between neutrons and protons which is the elusive deuteron-like type. In other words, all spin-triplet pairing correlations discussed in this work are also neutron-proton pairing correlations, that is, the most exotic type. While in principle spin-triplet pairing correlations can form between like-particles as well, they are expected to be less dominant than the ones considered here since they are suppressed by Pauli's exclusion principle. The second projection operator $P_{J_z=0}$ defines the spatial structure of the pairs which consist of particles paired in time-reversed axially-symmetric states with vanishing $z$-projection of the pair angular momentum $J$. Together, operators $P_{J_z=0}$ and $P_\alpha$ define the pairing scheme in which each pair forms in channel $\alpha$ with a wavefunction:
\begin{align}
\phi_{\alpha,pq}(\mfr_1,\mfr_2) &= \bra{pq} P_{J_z=0} P_{\alpha} \ket{\mf{r}_1\mf{r}_2} \notag \\ 
&= A_{\alpha,pq} \bracket{pq}{\mfr_1\mfr_2}~,\label{eq:pairwf}
\end{align}
where the channel amplitude $A$ is expressed in terms of the Clebsh-Gordan coefficients as
\begin{align}
A_{\alpha,ij}=  \delta_{j_{z_i} -j_{z_j}}   C^{\frac{1}{2}\frac{1}{2}S_{\alpha}}_{s_{z_i} s_{z_j} S_{z_\alpha}} C^{\frac{1}{2}\frac{1}{2}T_{\alpha}}_{t_{z_i}t_{z_j}T_{z_\alpha}}~.
\end{align}
It is necessary to always regulate zero-range interactions, to ensure sensible results. We do so by applying a window-like regulator in the single-particle energies. This means that only particles in states within a finite energy window feel the pairing interaction. We center this window roughly around the Fermi surface of the nuclei studied as this is where pairing correlations form. The window's width is to be determined in conjunction with the parameters $v_\alpha$ as the latter are strongly dependent on this regularization scheme (see tuning of the interaction in sec.~\ref{sec:tune}). The two-body interaction is expressed on the same  basis as the one-body Hamiltonian, as an interaction matrix $v_{ijpq}=\bra{ij}v(\mfr,\mfr')\ket{pq}$, and then antisymmetrized resulting in
\begin{align}
\bar{v}_{ij,pq} &= v_{ijpq}-v_{ijqp} \\
&=\sum_\alpha \frac{v_\alpha}{2\pi} \bar{A}_{\alpha,ij}\bar{A}_{\alpha,pq}\int \frac{d\rho dz}{\rho}  u^*_i u^*_j u_p u_q~, \label{eq:hfb_twobody}
\end{align}
where it is now written in terms of the anti-symmetrized channel amplitudes:
\begin{align}
\bar{A}_{\alpha,ij}=\sqrt{2}\delta_{j_{z_i} -j_{z_j}}   C^{\frac{1}{2}\frac{1}{2}S_{\alpha}}_{s_{z_i} s_{z_j} S_{z_\alpha}} C^{\frac{1}{2}\frac{1}{2}T_{\alpha}}_{t_{z_i}t_{z_j}T_{z_\alpha}}~.\label{eq:antiamps}
\end{align}

An HFB state is uniquely defined by the Bogolyubov transformation in Eq.~(\ref{eq:quasibasis}). Specifically, the transformation matrices $U$ and $V$ define the normal and anomalous densities,
\begin{align}
    \rho_{ij}&=\left(V^\star V^T\right)_{ij}~,\\
    \kappa_{ij}&=\left(U^\star V^T\right)_{ij}~,
\end{align}
and those in turn define the $\Gamma$ and $\Delta$ matrices:
\begin{align}
    \Gamma_{ij} &= \sum_{pq}\bar{v}_{iq,jp} \rho_{pq}~, \label{eq:hfb_gamma} \\
    \Delta_{ij} &= \frac{1}{2}\sum_{pq}\bar{v}_{ij,pq} \kappa_{pq}~. \label{eq:hfb_delta}
\end{align}
These matrices, uniquely define the HFB state, just like the matrices $U$ and $V$ do. For example, the expectation value of the Hamiltonian in Eq.~(\ref{eq:ham}), at an HFB state defined by the transformation matrices $U$ and $V$ (or the densities $\rho$ and $\kappa$, or even yet their corresponding $\Gamma$ and $\Delta$ matrices) is
\begin{align}    \bra{\textrm{HFB}}H\ket{\textrm{HFB}}=\textrm{Tr}\left[(\epsilon+\Gamma) \rho-\frac{1}{2}\Delta \kappa^\star\right] ~.
\end{align}
It is seen then that $\Gamma$ is simply a correction to the one-body term induced by the interaction and holds little pairing information. Furthermore, our interaction is a purely phenomenological one, regularized and designed to have the right properties, and it is eventually fit to prexisting phenomenology. As such one can neglect the correction $\Gamma$ and see it as included in the parameters of the single-particle potential. We discuss more details about this fitting in sec.~\ref{sec:tune}.

\subsection{The pairing amplitudes of different channels}
\label{sec:pairamp}
The $\Delta$ matrix quantifies the amplitude of the pairing correlations, but we want to be able to tell how these correlations are distributed across the six pairing channels. To that end, we will define pairing amplitudes for each channel separately. First define the pair field in the channel $\alpha$, using the pair wavefunction from Eq.~(\ref{eq:pairwf}), as
\begin{align}
\Pi_\alpha(\mfr,\mfr') = \sum_{ij} \phi_{\alpha,ij}(\mf{r},\mf{r}') = \bra{\mfr\mfr'}g_\alpha^\+ \ket{0} ~,
\end{align}
where the pair operator in the channel $\alpha$, namely  $g_\alpha$, is
\begin{align}
g_\alpha^\+ = \sum_{ij} P_\alpha P_x c_i^\+c_j^\+ = \sum_{ij} \bar{A}_{\alpha,ij}c_i^\+ c_j^\+ = c^\+ \bar{A}_\alpha c^\+~, ~\label{eq:g}
\end{align}
in terms of the antisymmetrized channel amplitudes $\bar{A}_{\alpha,ij}$from Eq.~(\ref{eq:antiamps}).
For computational convenience we can also define the hermitian pair-field operator (associated to the real part of the complex pair field $\Pi_\alpha$) to be
\begin{align}
G_\alpha= \frac{1}{2} \left(g_\alpha^\+ + g_\alpha\right) = \frac{1}{2}\left(c^\+\bar{A}_\alpha c^\+ - c \bar{A}^*_\alpha c\right) \label{eq:G}~.
\end{align}
 Note that $G_\alpha$ is hermitian by construction and its expectation value at the HFB ground-state takes a simple form:
\begin{align}
K_\alpha = \bra{\textrm{HFB}} G_\alpha \ket{\textrm{HFB}} = -\frac{1}{2}\left[\textrm{Tr}\left(\bar{A}_\alpha \kappa^*\right)+ \textrm{Tr}\left(\bar{A}_\alpha^*\kappa\right)\right] ~.
\end{align}
The quantities $K_\alpha$ are the pairing amplitudes in channel $\alpha$ and we use them to identify the symmetry properties (spin-singlet, spin-triplet, or mixed-spin) of an HFB state's pairing correlations. For instance, a state with $K_0=K_1=K_2=0$ has spin-triplet pair correlations. We don't differentiate between the different $S_z$ (or $T_z$) quantum numbers for each channel so we define normalized spin-singlet and spin-triplet amplitudes,
\begin{align}
x_S &= \frac{\left(\sum_{\alpha=1,2,3}K_\alpha^2\right)}{K^2}~, \label{eq:xs}\\
K^2&= \left(\sum_{\alpha=1}^6 K_\alpha^2\right)^{1/2}~.
\end{align} 
With these we can characterize the symmetry properties of a given HFB state:
\begin{align}
x_S^2 &\ge 4/5 \quad \textrm{singlet} \notag\\
1/5 <x_S^2 &< 4/5 \quad \textrm{mixed} \notag\\
x_S^2 & \le 1/5  \quad \textrm{triplet} \notag
\end{align}
When looking at ground states, the normalized amplitude is a function of the nucleus' proton and neutron numbers, and the deformation parametrizing its surface: $x_S=x_S(Z,N;\{\beta\})$. Then, by association, so are the symmetry properties of the pairing correlations. However, given the simple form of Eq.~(\ref{eq:G}), we can constrain the pairing properties of HFB states, and their symmetry, by constraining the expectation values $K_\alpha$. And so, for a given nucleus, we can identify its HFB ground state, i.e., the HFB state with the lowest energy, and its HFB states with purely spin-singlet and purely spin-triplet pairing symmetries.

\subsection{The block structure of the HFB pairing matrix}
\label{app:block-structure}
Owing to the symmetries of the one- and two-body terms employed in our HFB formulation, the standard HFB matrices take a block structure that greatly simplifies the calculations. The pairing matrix, defined in Eq.~(\ref{eq:hfb_delta}) is separated in blocks defined by the $j_z$ quantum number of the single particle states.

To demonstrate, assume the same numbering of states $i=(j_{zi},n_i,s_{zi},t_{zi})$ with the last quantum number ($t_{zi}$) changing the fastest, the second-to-last ($s_{zi}$) the second fastest, etc. For a shell of $N_{j_z}=2$ and $N_n=3$ proton and neutron states, the paring matrix is:
\begin{align}
\Delta &=\textrm{diag}(\Delta_{j_{z_1}},\Delta_{j_{z_2}})~, \\
    \Delta_{j_z} &= \left(\begin{array}{ccc}
            \Delta(n_3,n_3) & \Delta(n_3,n_2) & \Delta(n_3,n_1)   \\
            \Delta(n_2,n_3) & \Delta(n_2,n_2) & \Delta(n_2,n_1)   \\
             \Delta(n_1,n_3) & \Delta(n_1,n_2) & \Delta(n_1,n_1)   \\
        \end{array}\right)~,
\end{align}
suppressing the $j_z$ label in the matrices $\Delta(n_i,n_j)$. These, for $|j_z|<j_{z\,\textrm{max}}$, are:
\begin{align}
        &\Delta(n_i,n_j) =
        \left(\begin{array}{cc}
        \Delta_{\uparrow\uparrow,n_in_j} & \Delta_{\uparrow\downarrow,n_in_j} \\
        \Delta_{\downarrow\uparrow,n_in_j} & \Delta_{\downarrow\downarrow,n_in_j} 
        \end{array}\right) \\
        &=
         \left(\begin{array}{cccc}
        \Delta^{pp}_{\uparrow \uparrow, n_in_j}     & \Delta^{pn}_{\uparrow \uparrow, n_in_j} & \Delta^{pp}_{\uparrow \downarrow, n_in_j}  & \Delta^{pn}_{\uparrow \downarrow, n_in_j} \\
        \Delta^{np}_{\uparrow \uparrow, n_in_j}     & \Delta^{nn}_{\uparrow \uparrow, n_in_j} & \Delta^{np}_{\uparrow \downarrow, n_in_j}  & \Delta^{nn}_{\uparrow \downarrow, n_in_j}  \\
         \Delta^{pp}_{\downarrow \uparrow, n_in_j}     & \Delta^{pn}_{\downarrow \uparrow, n_in_j} & \Delta^{pp}_{\downarrow \downarrow, n_in_j}  & \Delta^{pn}_{\downarrow \downarrow, n_in_j} \\
         \Delta^{np}_{\downarrow \uparrow, n_in_j}     & \Delta^{nn}_{\downarrow \uparrow, n_in_j} & \Delta^{np}_{\downarrow \downarrow, n_in_j}  & \Delta^{nn}_{\downarrow \downarrow, n_in_j} 
\end{array}\right) ~,
\end{align}
and for $|j_z|=j_{z\,\textrm{max}}$ they are:
\begin{align}
\Delta(n_i,n_j) &=
        \Delta_{\uparrow\uparrow,n_in_j} =    \left(\begin{array}{cc}
        \Delta^{pp}_{\uparrow \uparrow, n_in_j}     & \Delta^{pn}_{\uparrow \uparrow, n_in_j} \\
        \Delta^{np}_{\uparrow \uparrow, n_in_j}     & \Delta^{nn}_{\uparrow \uparrow, n_in_j}
\end{array}\right)~, \\ 
\Delta(n_i,n_j) &=
        \Delta_{\downarrow\downarrow,n_in_j} =    \left(\begin{array}{cc}
        \Delta^{pp}_{\downarrow \downarrow, n_in_j}     & \Delta^{pn}_{\downarrow \downarrow, n_in_j} \\
        \Delta^{np}_{\downarrow \downarrow, n_in_j}     & \Delta^{nn}_{\downarrow \downarrow, n_in_j}
\end{array}\right)~,
\end{align}
for $j_z=j_{z\,\textrm{max}}$ or $j_z=-j_{z\,\textrm{max}}$, respectively.
In general, there are $2N_{j_z}$ blocks, each block signified by $\Delta_{j_z}$ is a $4N_n\times 4N_n $ matrix, except for the two blocks that correspond to $j_z=|j_{z\,\textrm{max}}|$ which have dimensions $2N_n\times 2N_n$. 

\section{Spin symmetries of pairing in deformed nuclei}
\label{sec:study}
The formulation of HFB theory that was described above was developed to address nuclear deformation. Specifically, we want to see the interplay between deformation and the competition of spin-singlet and spin-triplet pairing. We focus on the light Lanthanides. These are nuclei close to the $N=Z$ line with mass number $A\sim 130$ and they are the prime candidates for substantial pairing correlations with various spin-symmetries~\cite{bertsch:2011,gezerlis:2011}. These are to be compared with the vast majority of nuclei that display only spin-singlet correlations in their ground states. Nevertheless, the results of our study of the interplay between singlet pairing and deformation can be extrapolated to other parts of the nuclear chart as well. In the rest of this section we define the main quantities that we will use to probe the interplay of deformation and different spin-symmetries of pairing correlations, we tune the zero-range interaction shown in Sec.~\ref{sec:pairinter} to be applicable to the Lanthanides region, and finally we perform a detailed study of the effect of deformation on pairing correlations and their spin-symmetry.

\subsection{Pairing amplitudes, correlation energies, and the odd-even staggering}
For each HFB state, the ``tag'' of a specific spin-symmetry, i.e., specific parity under particle exchange, was defined in detail in Sec.~\ref{sec:pairamp}. Since those are normalized ($0\le x_S\le 1$), they do not inform us about the strength of the pairing correlations but only the relative strength of the spin-singlet channel compared to the spin-triplet one. For a measure of the pairing's strength in the many-body wavefunction  we use the pairing correlation energy defined below.

Quantifying the strength of the pairing correlations is necessary because the pairing amplitudes do not tell the whole story: an HFB state could be nearly degenerate with the normal state which would hint towards vanishing pairing correlations. On the other extreme, two or more pairing states of different spin symmetries could be nearly degenerate while well separated in energy from the normal state, hinting on an instability. In fewer words, the energy difference between different pairing states is as essential to their analysis as their pairing amplitudes. The easiest way to quantify the strength of the pairing correlation is to measure the energy of an HFB state from some reference state~\cite{frauendorf:2014}. We define our reference state as the normal state, i.e., the HFB state with vanishing pairing amplitudes. In practice, we can identify this state by constraining all pairing amplitudes to $0$ during the gradient descent. Naming the energy of the reference state $E_0$, this correlation energy is:
\begin{align}
E_{\rm corr} = E-E_0~. \label{eq:corr}
\end{align}
When looking at ground states, the correlation energy of a nucleus will depend on its numbers of protons and neutrons, and its deformation: $E_{\rm corr} = E_{\rm corr}(Z,N;\{\beta\})$, similarly to $x_S(Z,N;\{\beta\})$. Note that even though the correlation energy is well defined, its connection to experiment is ambiguous and it can be best described as an increase in binding energy due to the pairing correlations. Hence, we also calculate the traditional signature of pairing correlations and measure of their strength: the pairing gap defined via an odd-even staggering~\cite{frauendorf:2014,duguet:2001}:
\begin{align}
    \Delta(n) = E_{\textrm{corr}}(n)-\frac{1}{2}\left[E_{\textrm{corr}}(n-1)+E_{\textrm{corr}}(n+1)\right]~,\label{eq:oes}
\end{align}
where $n$ is an odd number of one nuclear species while the other is kept at an even number. This can provide a measure of pairing correlations that is accessible in the lab via precision mass measurements, but when calculated it should always be compared to the mean single-particle level spacing, as that can also cause a similar effect.

Given the above discussion we investigate the interplay between spin-singlet and spin-triplet pairing, and deformation by calculating pairing amplitudes and correlation energies. Note that writing  $x_S(Z,N;\{\beta\})$ and $E(Z,N;\{\beta\})$ is implying that the deformation parameters $\{\beta\}$ are independent of the neutron and proton numbers or that a single nucleus can have different shapes. While nuclei are known to change shapes when excited (see \textit{shape coexistence}), our motivation for keeping deformation, and proton and neutron numbers decoupled is different. The ground state of nuclei has a uniquely defined shape identified via some shape moment which can be connected to measurable quantities, e.g., electromagnetic moments. Based on such measurements the dependence of the deformation parameters on neutron and proton numbers is well-defined and available for light- to medium-mass nuclei. For heavy isotopes, such as the ones studied here, these measurements are scarce and so, even though we will often turn to the tabulation in Ref.~\cite{ref:moller} for a prediction of the realistic deformation, we will still explore a number of possible deformations for the nuclei we will study. Furthermore, in a mean-field approach such as HFB, one must include deformation by hand by breaking the rotational invariance of the Hamiltonian. That is, in our description, deformation is not emergent and comes as prior knowledge, predicted and prescribed, and so it should not be fine-tuned.

\subsection{Tuning a contact interaction for the Lanthanides region}
\label{sec:tune}

A study of the pairing correlations in the Lanthanides region was performed in Ref.~\cite{gezerlis:2011} based on the HFB approach formulated in Ref.~\cite{bertsch:2011} where deformation was not taken into account. With that being the only available literature for pairing correlations of different spin-symmetries in this region, we wish to have it as a point of departure: we tune our interaction to reproduce the results of Refs.~\cite{gezerlis:2011} and \cite{bertsch:2011} at the absence of deformation. 

While the interaction in Refs.~\cite{gezerlis:2011} and \cite{bertsch:2011} is also of zero-range, it contains a different operator structure, consistent with the type of pairs that one expects in the ground-state of a spherical nucleus. There only the $L=0$ parts of the two-particle wavefunctions interact, yielding pairs of $J=0$ or $J=1$, and interaction strengths in the singlet and triplet channels are set to $v_s=300$ MeV and $v_t=450$ MeV by fitting to phenomenological shell-model Hamiltonians for $sd$- and $fp$-shell nuclei. Note that this corresponds to $v_t/v_s\approx1.5$, as is the case for the two channels in most realistic effective interactions. Note also that the interaction strengths in all spin-singlet channels are set equal and similarly is the case of the spin-triplet channels; we keep this feature. We also keep the $10~\textrm{MeV}$ width of the regulating window for the contact interaction, centered at $-13~\textrm{MeV}$. Our contact interaction has a more general operator structure restricting only the $z$-projection of the interacting parts in the pair wavefunctions to be the $J_z=0$ [see Eq. (\ref{eq:pairint})]. This is again inspired by the type of pairing correlations expected in the ground state of a deformed nucleus where particles could pair in states of $J>1$. While one would still anticipate a similar ratio of $v_t/v_s$, the absolute values of the couplings can be different. We find that $v_s=87~\textrm{MeV}$ and $v_t=120~\textrm{MeV}$ best describe the phenomenology of Refs.~\cite{bertsch:2011}~and~\cite{gezerlis:2011}; we describe the tuning process below.

\begin{figure}
\includegraphics[width=0.45\textwidth]{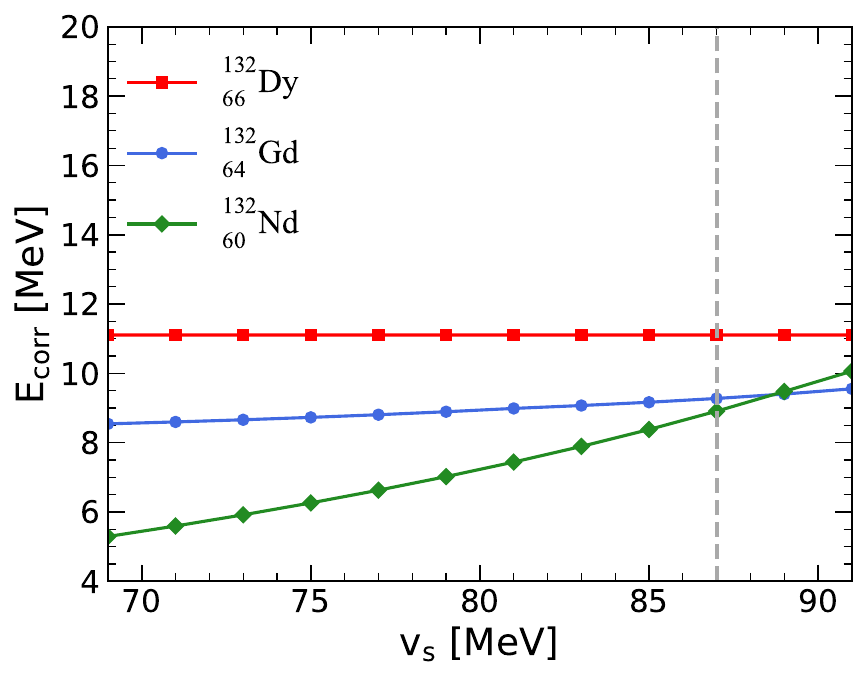}
\caption{The correlation energies of the selected nuclei as a function of the interaction strength in the singlet channel, $v_s$. The strength in the triplet channel is kept fixed to $v_t=120~\textrm{MeV}$. The gray vertical dashed line corresponds to the value of $v_s$ chosen.}
\label{fig:vs_e}
\end{figure}

\begin{figure}
\includegraphics[width=0.45\textwidth]{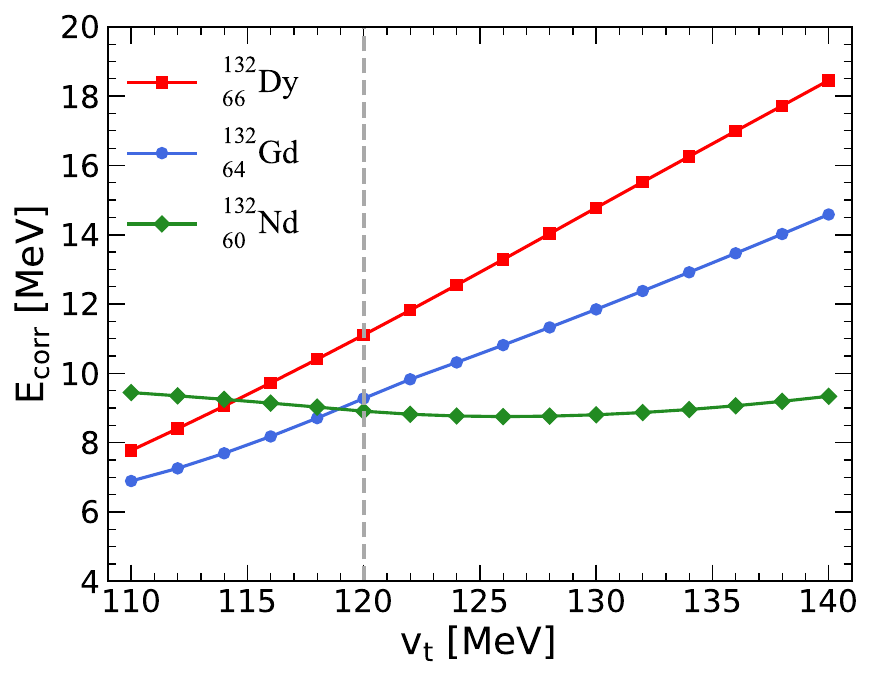}
\caption{The correlation energies of the selected nuclei as a function of the interaction strength in the triplet channel, $v_t$. The strength in the singlet channel is kept fixed to $v_s=87~\textrm{MeV}$.The gray vertical dashed line corresponds to the value of $v_t$ chosen.}
\label{fig:vt_e}
\end{figure}

We focus on the $A=132$ isobar and first on three nuclei therein, \dy, \gd, and \nd, with dominant spin-triplet, mixed-spin, and spin-singlet pairing correlations in their ground states, respectively. Figures \ref{fig:vs_e} and \ref{fig:vt_e} show their ground state pairing amplitudes and correlation energies as a function of $v_s$, for constant $v_t=120~\textrm{MeV}$, and as a function of $v_t$ for constant $v_s=87~\textrm{MeV}$, respectively.

These calculations also serve as sanity-checks of our phenomenology. In Fig.~\ref{fig:vs_e}, for constant interaction strength in the triplet channel, the correlation energy of \dy stays constant reaffirming that its ground state is dominated by spin-triplet pairing correlations. In the same figure, the correlation energy of the other two nuclei, which have at least some spin-singlet component in their pairing correlations, rises linearly with increasing $v_s$, with slopes telling of the dominance of spin-singlet pairing. Figure~\ref{fig:vt_e} depicts similar behaviors when varying the interaction strength in the spin-singlet channel: an approximately constant correlation energy for \nd, whose ground state has predominantly spin-singlet pairing correlations, and a rise in correlation energies for the other two nuclei with increasing $v_t$. As with Fig.~\ref{fig:vs_e}, the correlation energy of the nucleus with the mixed-spin pairing rises with the smaller slope.

\begin{figure}
\includegraphics[width=0.45\textwidth]{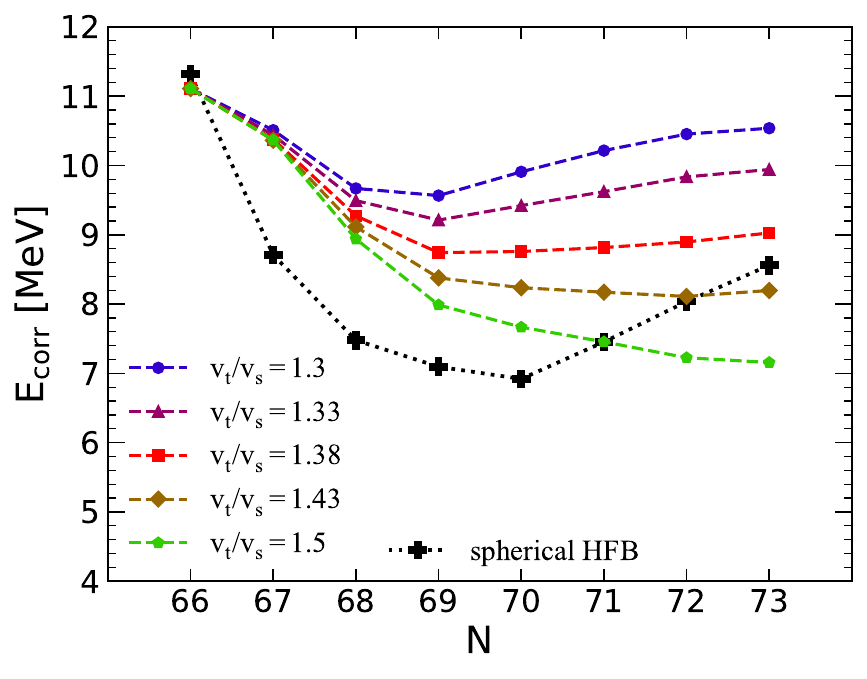}
\caption{The correlation energies on the $A=132$ isobar for a range of ratios of interaction strengths $v_t/v_s$. The black crosses show the same quantity in the spherical HFB model of Refs.~\cite{bertsch:2011,gezerlis:2011}. Lines are drawn to guide the eye.}
\label{fig:nz_e}
\end{figure}

\begin{figure}
\includegraphics[width=0.45\textwidth]{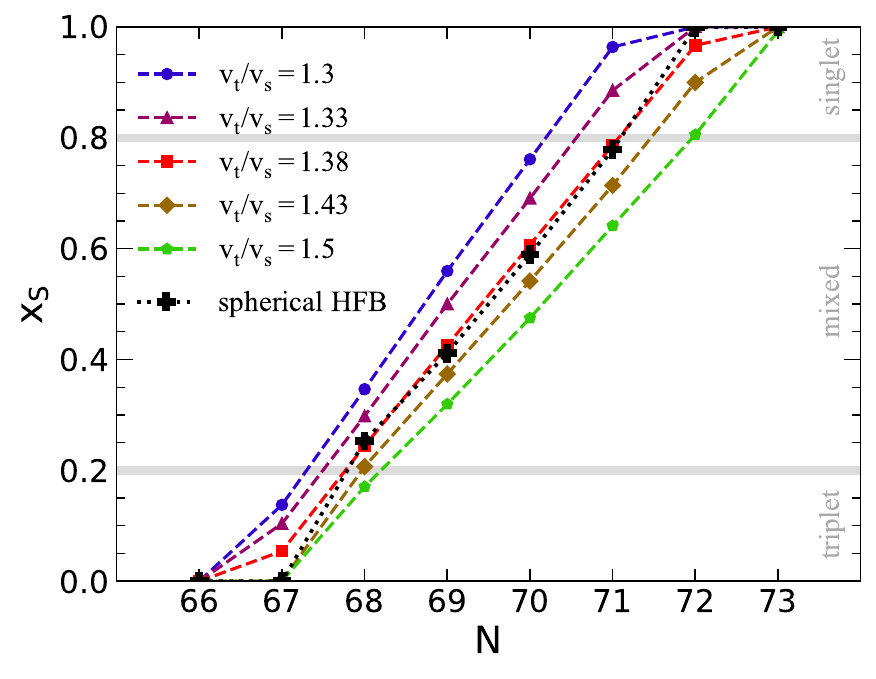}
\caption{The pairing amplitudes on the $A=132$ isobar, as defined in Eq.~(\ref{eq:xs}). The black crosses show the same quantity in the spherical HFB model of Refs.~\cite{bertsch:2011,gezerlis:2011}. Lines are drawn to guide the eye.}
\label{fig:nz_xs}
\end{figure}

From the linear behavior of \dy's correlation energy in Fig.~\ref{fig:vt_e} we can choose the tuning in the spin-triplet channel to $v_t=120~\textrm{MeV}$ as this best reproduces this nucleus' correlation energy from Ref.~\cite{gezerlis:2011}. Then, keeping $v_t$ constant we vary $v_s$ looking at other nuclei with various isospin asymmetries on the $A=132$ isobar. We plot their correlation energies in Fig.~\ref{fig:nz_e} and their pairing amplitudes in Fig.~\ref{fig:nz_xs} where we keep $v_t/v_s$ around $1.5$ anticipating that ratio, as mentioned before. The correlation energy in the spherical limit of Refs~\cite{bertsch:2011,gezerlis:2011} reaches a local minimum at $N=70$ and the ratio $v_t/v_s$ in our formulation controls the depth and width of this dip. The pairing amplitudes display a mostly linear behavior both for the spherical HFB and our formulation and the two can be easily matched by the right $v_t/v_s$ ratio. Combining these together, we find that $v_t/v_s\approx 1.38$, or $v_s=87~\textrm{MeV}$, best reproduces the features of the spherical HFB calculations for both quantities. 

\subsection{The pairing's spin-symmetry and deformation} 

Detailed studies using the Finite Range Liquid Drop Model (FRDM) show that the region of $A\sim 130$ close to the $N=Z$ line is characterized by $\beta_2\sim 0.25$, relatively uniformly, and finite but small $\beta_4$ and $\beta_6$ deformation ~\cite{ref:moller}. Note that a comparison between the $\beta_\lambda$ parameters from Ref.~\cite{ref:moller}'s standard tabulation and those of the Cassini-oval formulation is only possible when keeping $\epsilon=0$ (see sec.~\ref{sec:deform}). As discussed in sec.~\ref{sec:deform}, here deformation enters via the one-body potential that defines the single-particle states available to the nucleons and so its effect enters only by modifying the single-particle wavefunctions. Based on the predictions of Ref.~\cite{ref:moller}, quadrupole deformation seemingly dominates with a relatively uniform value in that area of the nuclear chart. Hence we start with a detailed study of the effect that this type of deformation has on the pairing correlations. From the lessons learned, we'll then turn to higher-modes of deformation.

\subsubsection{Quadrupole deformation}
\label{sec:beta_2}  

\begin{figure}
\includegraphics[width=0.45\textwidth]{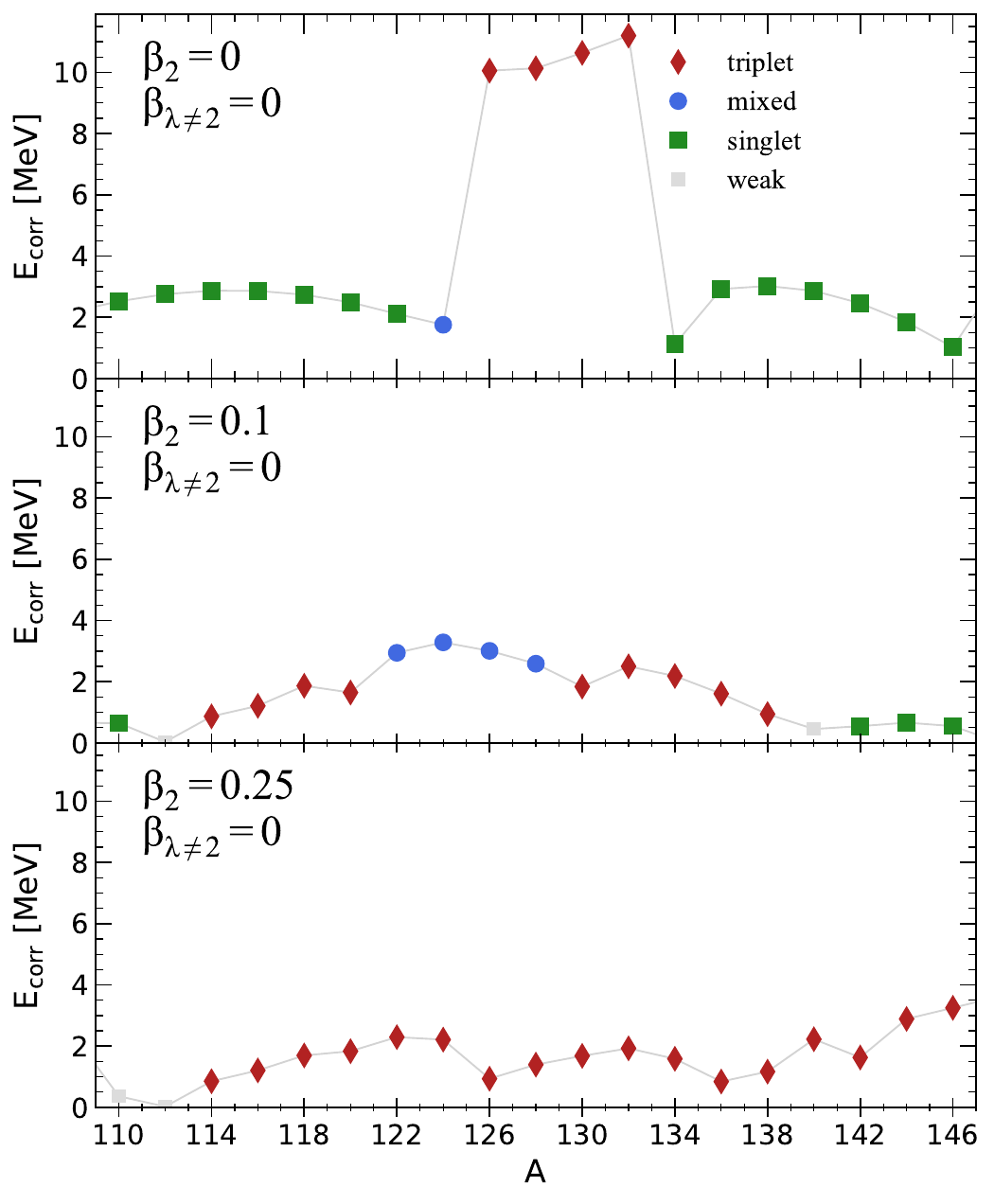}
\caption{The correlation energies of nuclei on the $N=Z$ line and their changes with deformation. From top to bottom: we plot the correlation energies with no deformation ($\beta_\lambda=0$), small $\beta_2$ deformation ($\beta_2=0.1,\,\beta_{\lambda\neq2}=0$), and the average $\beta_2$ deformation in the region $\beta_2=0.25,\,\beta_{\lambda\neq2}=0$. The solid lines are included to guide the eye.}
\label{fig:nz0_b2all}
\end{figure}

Quadrupole deformation which is parametrized by the $\beta_2$ parameter in Eq.~(\ref{eq:std_expan}) is the simplest non-trivial deformation one can start from because in an expansion of the form in Eq.~(\ref{eq:std_expan}), $\beta_0$ is fixed by the volume of the nucleus and $\beta_1$ corresponds mainly (for small deformations) to a translation of the whole system (see, e.g., Ref.~\cite{book:NuclMBP}). We first look at the $N=Z$ line at $A\sim 130$. Figure~\ref{fig:nz0_b2all} shows the evolution of the pairing correlations from no deformation (top panel), to small deformation, $\beta_2=0.1$ (middle panel), to $\beta_2=0.25$, i.e., the average $\beta_2$ in the region (bottom panel). The $y$-axis measures the correlation energy for each nucleus' ground state while red diamonds correspond to spin-triplet pairing, blue circles to mixed-spin, and green squares to spin-singlet; this color- and shape-coding is kept consistent throughout, unless specified otherwise. The chosen values of $\beta_2=0,0.1,0.25$ can be taken as slices of negligible, small, and moderate deformation, respectively, which we will now analyze in detail starting from their effect on the $N=Z$ line seen in Fig.~\ref{fig:nz0_b2all} and later move away from it to finite isospin asymmetry.

\paragraph{The line of $N=Z$}
At no deformation, we recover the features of Refs.~\cite{bertsch:2011} and \cite{gezerlis:2011}'s description: an increase in the correlation energies and an emergence of spin-triplet pairing in a region around $A=132$ where low-$l$ single-particle states lie close to the Fermi surface. As deformation increases, these single-particle states split according to their orbital angular momentum's $z$-projection, $l_z$, and the correlation energy of all types of pairing correlations are suppressed by a factor of $\sim 5$.

At relatively low deformation ($\beta_2=0.1$), where the structure of the single-particle levels is only lightly modified, we see a remnant of the peak in correlation energies found at the absence of deformation. Spin-triplet pairing dominates in the ground states of most nuclei of the $N=Z$ line in the region already at this deformation.

\begin{figure}
\includegraphics[width=0.5\textwidth]{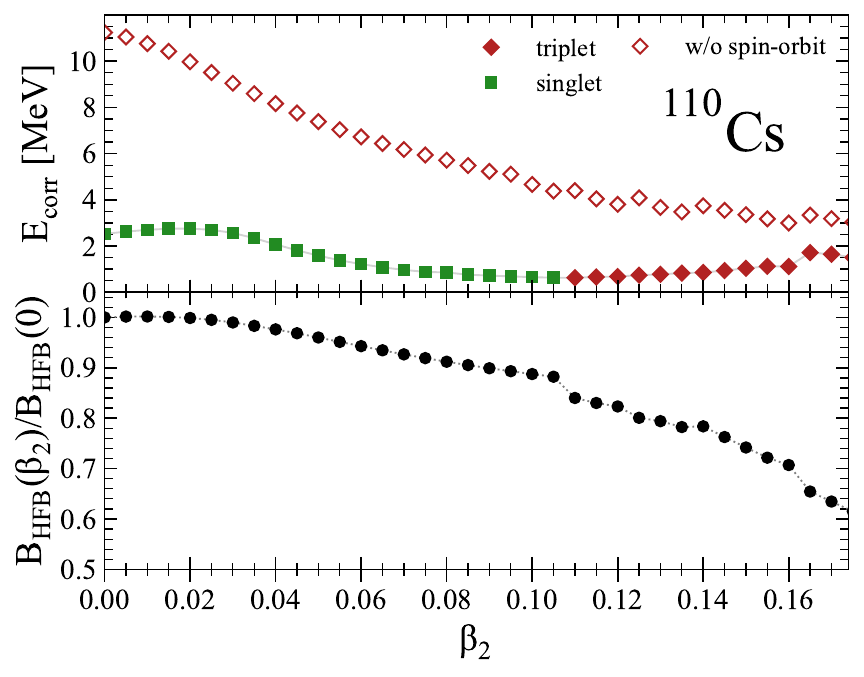}
\caption{The evolution of the correlation energies in the ground state of $^{110}$Cs that lies on the $N=Z$ line. This nucleus' ground state contains dominant spin-singlet pairing correlations at no deformation which quickly give way to spin-triplet once deformed.}
\label{fig:110cs}
\end{figure}

A typical example is the ground state of $^{110}$Cs whose pairing correlations can be seen in Fig.~\ref{fig:110cs} as they evolve with increasing $\beta_2$ deformation. Specifically, in the top panel of Fig.~\ref{fig:110cs} the y-axis measures the correlation energy and the colour- and shape-coding describes the dominant pairing correlations. We also include calculations where the spin-orbit field has been artificially turned off which we plot with open symbols. In the bottom panel, we plot the strength of the spin-orbit field as a fraction of its strength in the spherical limit. The $^{110}$Cs nucleus resides on the $N=Z$ line with purely spin-singlet pairing correlations in its ground state when assumed to be spherical. However, when deformation is turned on and closer to its expected physical value the  spin-singlet pairing correlations are suppressed by the quadrupole deformation and spin-triplet pairing correlations start to dominate at $\beta_2\sim 0.1$. In the bottom panel of Fig.~\ref{fig:110cs} the spin-orbit field is seen to drive this transition by decreasing and allowing spin-triplet correlations to form. This picture is reinforced when we find that if we turn off the spin-orbit field artificially, the nucleus' ground state exhibits spin-triplet pairing correlations for any $\beta_2$ value and the associated correlation energy is suppressed by deformation at a lower rate than the spin-singlet one.

\begin{figure}
\includegraphics[width=0.5\textwidth]{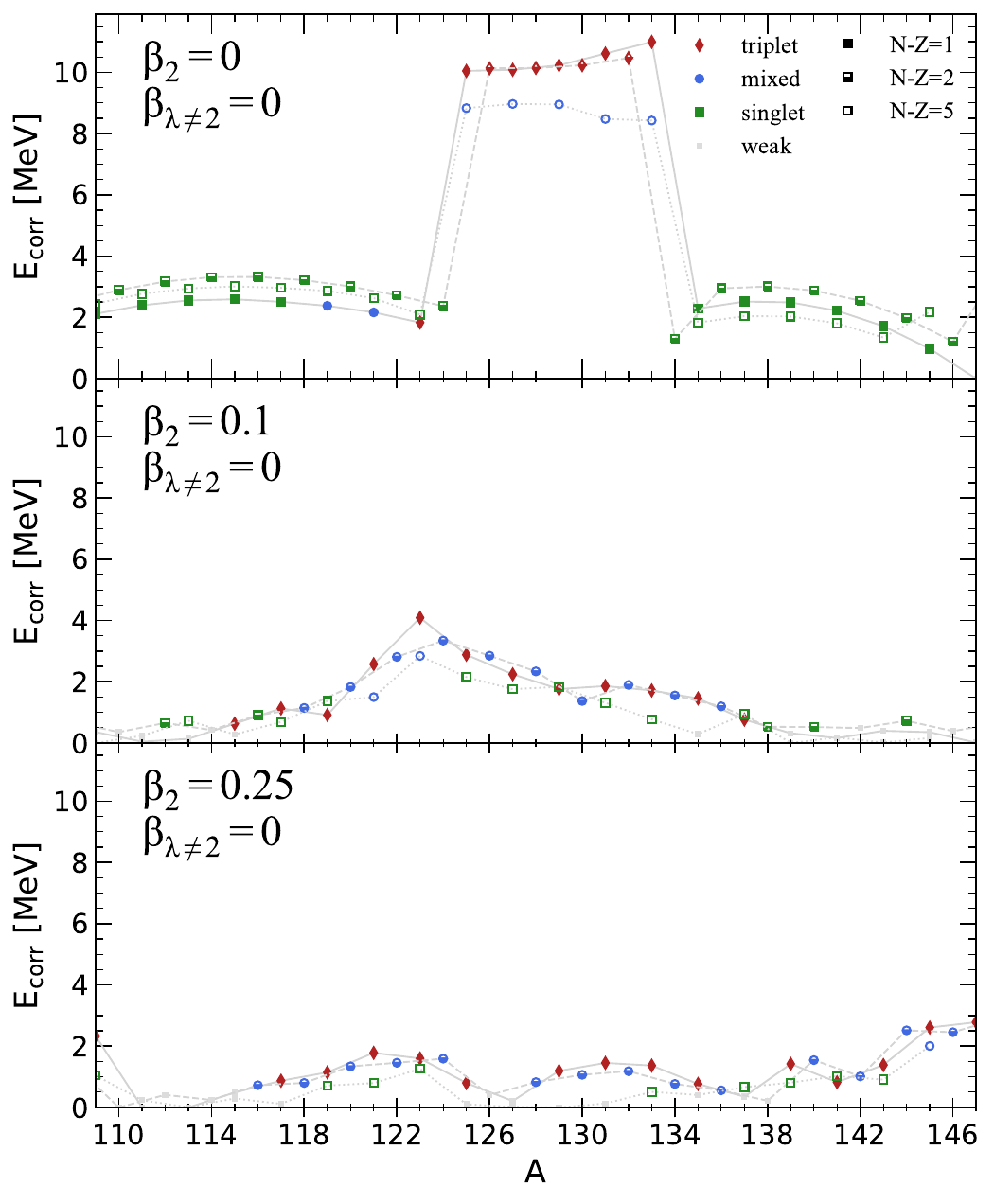}
\caption{The evolution of the correlation energies away from the $N=Z$ line. From left to right and top to bottom, we have set $\beta_2=0.01,~0.1,~0.25$.}
\label{fig:nz125_b2all}
\end{figure}

\begin{figure}
\includegraphics[width=0.5\textwidth]{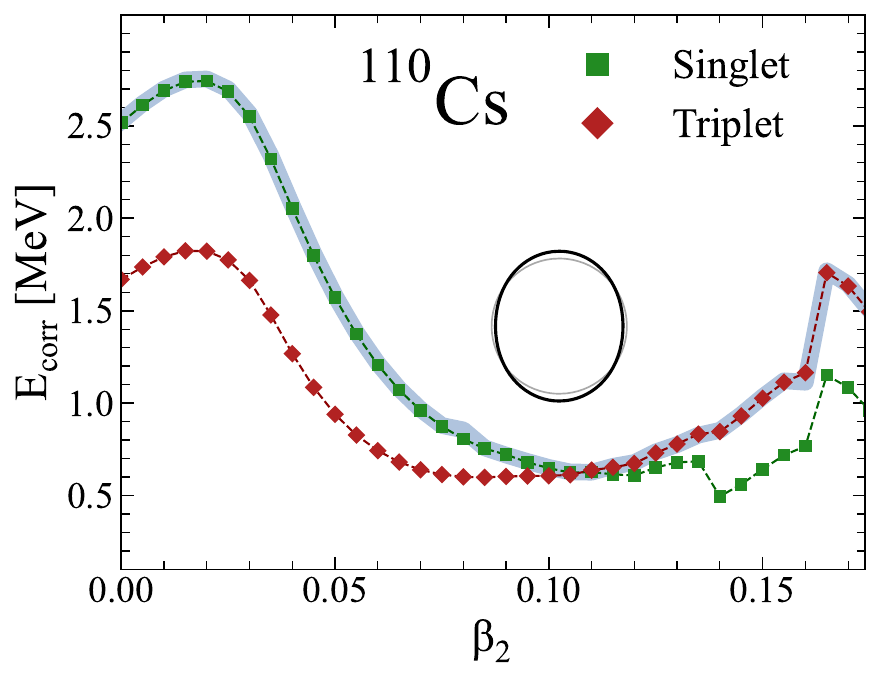}
\caption{The evolution of the correlation energies of two different HFB states describing $^{110}$Cs that are constrained to only spin-singlet (green squares) or spin-triplet (red-diamonds) pairing correlations. Dashed lines are included to guide the eye. The solid blue line corresponds to the unconstrained correlation energies shown in Fig.~\ref{fig:110cs}. The solid black line shows the shape of the nucleus at deformation $\beta_2=0.1$, where the two types of pairing correlations have equal correlation energies, while the nucleus' shape at the spherical limit is included in grey for comparison.}
\label{fig:110cs_constr}
\end{figure}

Figure~\ref{fig:110cs} suggests that spin-triplet pairing when dominant is mostly affected by the spin-orbit field, that is, the effect of deformation on spin-triplet pairing seems to be indirect at the mean-field level and mediated mainly through the spin-orbit field. This is consistent with our current picture of the spatial arrangement of the pairing correlations: when spin-triplet pairing dominates it forms in the interior of the nucleus while singlet pairing forms closer to the surface. This arrangement is once again caused by the larger effect of the spin-orbit field on particles orbiting close to the nuclear surface, which is detrimental to the spin-triplet pairing correlations. With spin-triplet pairing forming mainly in the interior, it is affected less by changes to the surface unless when it is forced outwards by the lack of low-$l$ single-particle states close to the Fermi surface. This effect is seen more clearly in Fig.~\ref{fig:110cs_constr} where the response of the two types of pairing correlations to quadrupole deformation is completely disentangled. There we plot two different types of HFB states for $^{110}$Cs: one where the pairing correlations are constrained to be spin-triplet (red diamonds) and one constrained to be spin-singlet pairing correlations (green squares); see sec.~\ref{sec:pairamp} for details on how these states are obtained. The correlation energies of these HFB states at any given deformation are plotted as a function of the quadrupole deformation. At a fixed deformation, the state with the highest correlation energy corresponds to the HFB ground state whose correlation energy is marked by a blue solid line. Again we see that the state with spin-triplet pairing correlations is affected less by deformation at low $\beta_2$ values, which again can be traced back to the only moderately modified spin-orbit field in Fig.~\ref{fig:110cs}. At $\beta_2\approx0.11$, the spin-triplet correlation energy overtakes that of the state with spin-singlet pairing, whose correlation energy is suppressed more in comparison and the ground state of $^{110}$Cs forms dominant spin-triplet pairing correlations.

\begin{figure}
\includegraphics[width=0.5\textwidth]{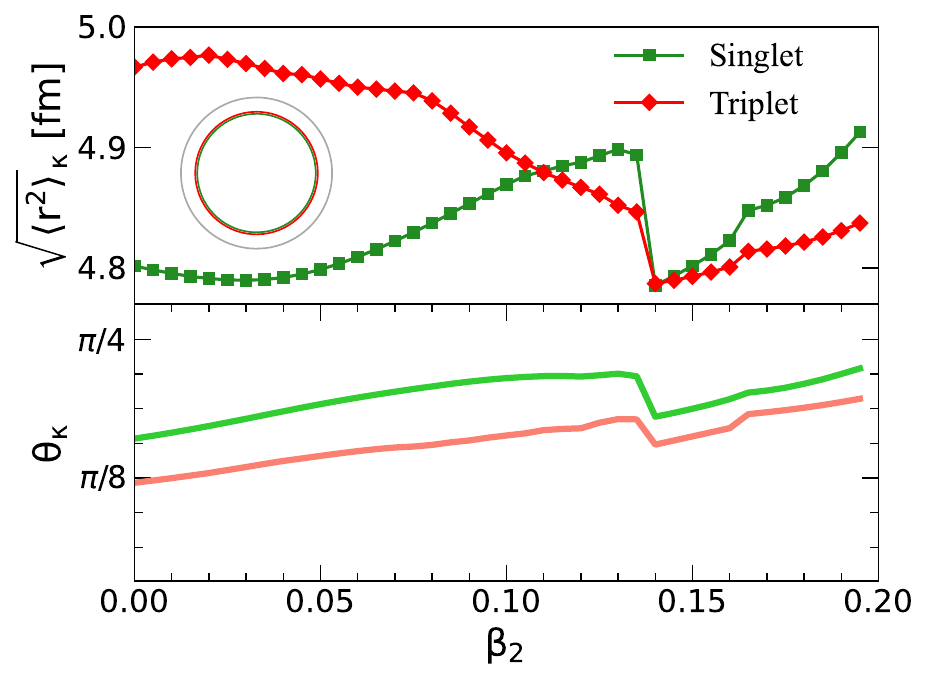}
\caption{The spatial configuration of the pairing correlations in $^{110}$Cs and their evolution with $\beta_2$ deformation. The top panel shows the root-mean-squared radius of the pairing fields in the spin-singlet and spin-triplet channel as defined in Eq.~(\ref{eq:rms}) In the inset of the top panel we demonstrate the circles that these radii define at the spherical limit ($\beta_2=0$) in a cross-section of the nucleus. The bottom panel shows the angular configuration of the pairing correlations; see Eq.~(\ref{eq:theta_kappa}) for its definition and the text for details}
\label{fig:110cs_rms}
\end{figure}

From the analysis of Figs.~\ref{fig:110cs} and \ref{fig:110cs_constr}, we understand that the spatial structure of the pairing correlations can play a role in their strength due to the presence of the strong spin-orbit field in the nucleus' surface. In an attempt to quantify this connection in the top panel Fig.~\ref{fig:110cs_rms} we plot the root-mean-squared radius of the pairing correlations in each of the two constrained states shown in Fig~\ref{fig:110cs_constr}, defined as,
\begin{align}
    \sqrt{\mv{r^2}_\kappa} = \left[\frac{\int d\rho dz (\rho^2+z^2) \kappa^2(\rho,z)}{\int d\rho dz  \kappa^2(\rho,z)}\right]^{1/2}~, \label{eq:rms}
\end{align}
where $\kappa$ corresponds to the pairing field in the dominant channel $\alpha$, 
\begin{align}
    \kappa_\alpha (\rho,z) = \frac{1}{4\pi} \sum _{ij} u_i(\rho,z) u_j(\rho,z) \bar{A}_{\alpha,ij} \kappa_{ij}.
\end{align}
Since $\kappa_\alpha$ can act as the wavefunction of the a condensates of pairs in the channel $\alpha$, the quantity $\sqrt{\mv{r^2}_\kappa}$ can be interpreted as the root-mean-squared radius of a pair in the most dominant pairing channel. We find that at the spherical limit, spin-triplet pair correlations happen closer to the nucleus' surface than spin-singlet ones; in that limit spin-triplet pairing does not form in the ground state which is dominated by spin-singlet. As quadrupole deformation increases, the spin-triplet pair correlations move towards the interior while the spin-singlet ones shift towards the surface. Once spin-triplet pairing gets closer to the nucleus' center than spin-singlet, it dominates and forms in the ground state. Note the radius of $^{110}$Cs is $\approx 6.12~\textrm{fm}$ and so the shifts observed in Fig.~\ref{fig:110cs_rms} are small. Nonetheless, the nucleus' ground state changes pairing channel exactly when the root-mean-squared radii of the two channels are equal. Finally, in the bottom panel of Fig.~\ref{fig:110cs_rms} we plot the angle of the pairing correlations,
\begin{align}
    \theta_{\kappa}=\cos^{-1}\left(\frac{\sqrt{\mv{z^2}_\kappa}}{\sqrt{\mv{r^2}_\kappa}}\right)~, \label{eq:theta_kappa}
\end{align}
where $\mv{z^2}_\kappa$ is defined similarly to $\mv{r^2}_\kappa$ in Eq.~(\ref{eq:rms}). The angle of the pairing correlations in the two channels tells of the three-dimensional arrangement of the correlations in the nucleus, with spin-triplet pairing being always further from the $z$-axis. This once again can be traced back to these correlations forming between particles in low-$l$ single-particle states and so lacking high-$l_z$ components.

\begin{figure}
\includegraphics[width=0.5\textwidth]{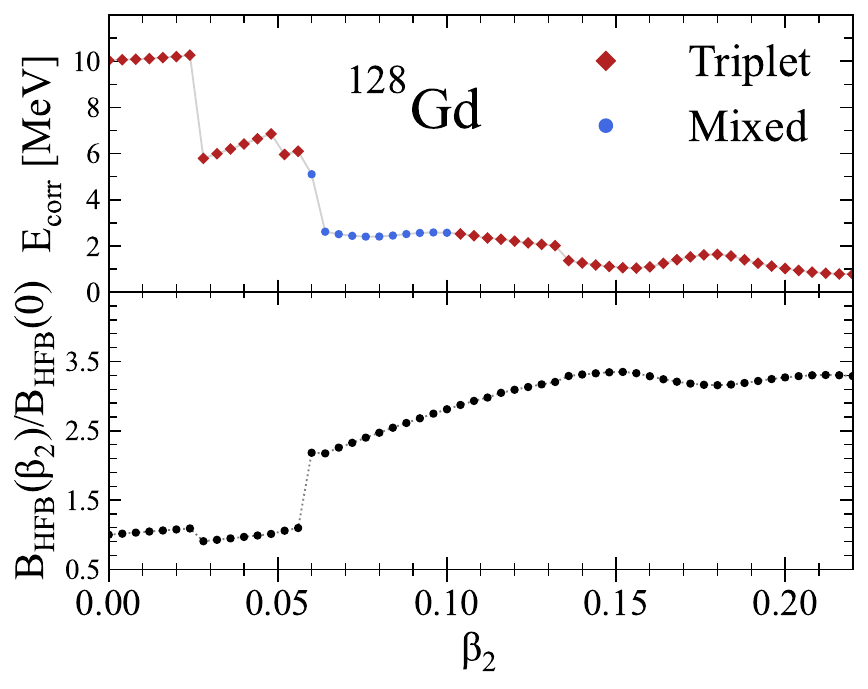}
\caption{The evolution of the correlation energies in the ground state of $^{128}$Gd that lies on the $N=Z$ line. This nucleus' ground state contains dominant spin-triplet pairing correlations at moderate as well as no quadrupole deformation, while at low deformation mixed-spin pairing correlations emerge.}
\label{fig:128gd}
\end{figure}

Going back to the systematics shown in Fig.~\ref{fig:nz0_b2all}, apart from nuclei like $^{110}$Cs that develop spin-triplet pairing correlations in their ground state at small deformation, we also find some nuclei, e.g., $^{128}$Gd, with spin-triplet pairing correlations already in their ground states at the spherical limit, that display mixed-spin pairing correlations at these small deformations. This is pointing to a weakening of the spin-triplet pairing correlations by $\beta_2$ which at first glance contradicts the conclusion from the study of $^{110}$Cs's ground state. A closer look in the evolution of $^{128}$Gd can be seen in Fig.~\ref{fig:128gd} where, in a manner similar to Fig.~\ref{fig:110cs}, in the top panel the y-axis measures the correlation energy of the ground state while the color- and shape-coding shows the corresponding spin-symmetry of the pairing correlations and the bottom panel shows the strength of the spin-orbit field, appropriately normalized. There it is clearly seen then that the weakening of the spin-triplet pairing, allowing the emergence of mixed-spin pairing, is caused by an amplification of the spin-orbit field that starts at $\beta_2\sim 0.05$. This in turn can be traced back to paired particles occupying single-particle states that lie closer to the nuclear surface as the energies of those states approach the Fermi surface. At deformation $\beta_2=0.1$, the spin-orbit field reaches a plateau becoming almost independent of the increasing deformation. However, with deformation increasing, the spin-singlet pairing correlations are weakened yielding to the spin-triplet ones, which as seen before mainly respond to the effect of the spin-orbit field. The adversarial effect of the spin-orbit field on the triplet-pairing correlations can be seen clearly here: the correlation energy has dips at the spin-orbit field's peaks. Specifically, at $\beta_2\approx0.15$ and $\beta_2\approx0.22$, two $l=4$ single-particle levels approach and cross the Fermi surface as they are shifted to lower energies by the deformation. Their transit is marked by the peaks in the otherwise flat spin-orbit field that induce the corresponding dips in the spin-triplet correlation energy. Note that the discontinuities are artifacts of our regulation of the contact interaction discussed in sec.~\ref{sec:pairinter}: single particle states that exit the regulation window suddenly drop their interaction energy which introduces artificial shell effects; these would be smoothed out but still qualitatively present in different regulation schemes.

\begin{figure}
\includegraphics[width=0.5\textwidth]{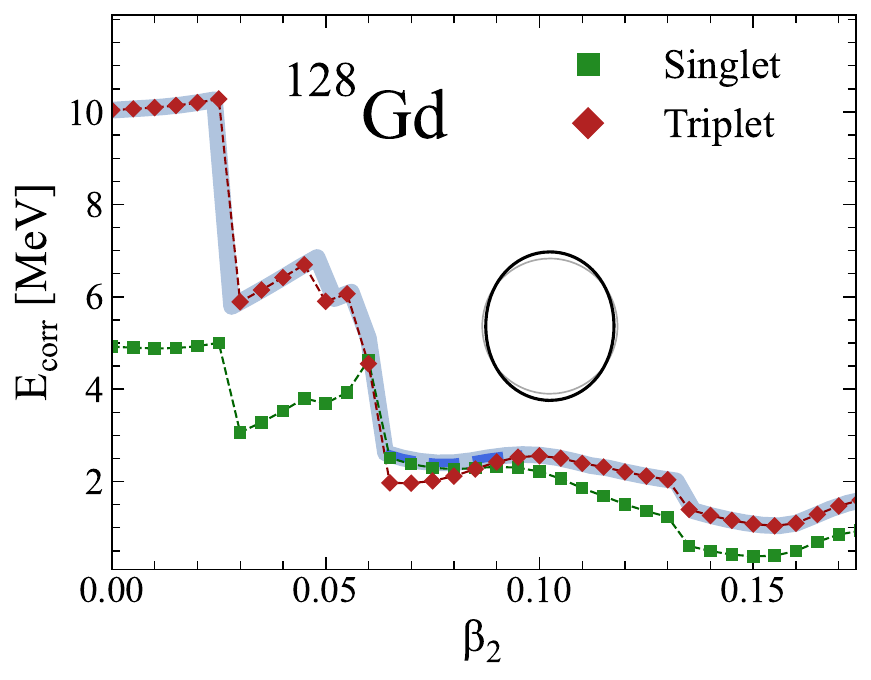}
\caption{The evolution of the correlation energies of two different HFB states describing $^{128}$Gd that are constrained to only spin-singlet (green squares) or spin-triplet (red-diamonds) pairing correlations. Dotted lines are included to guide the eye. The solid blue line corresponds to the unconstrained correlation energies shown in Fig.~\ref{fig:128gd} while the dashed blue line marks the area where mixed-spin pairing emerges. The emergence of mixed-spin pairing correlations seen in Fig.~\ref{fig:128gd} is seen here too when the spin-triplet correlation energy comes sufficiently close to the spin-singlet one.The solid black line shows the shape of the nucleus at deformation $\beta_2=0.11$, where the spin-triplet pairing correlations take over, while the nucleus' shape at the spherical limit is included in grey for comparison.}
\label{fig:128gd_constr}
\end{figure}

\begin{figure}
\includegraphics[width=0.5\textwidth]{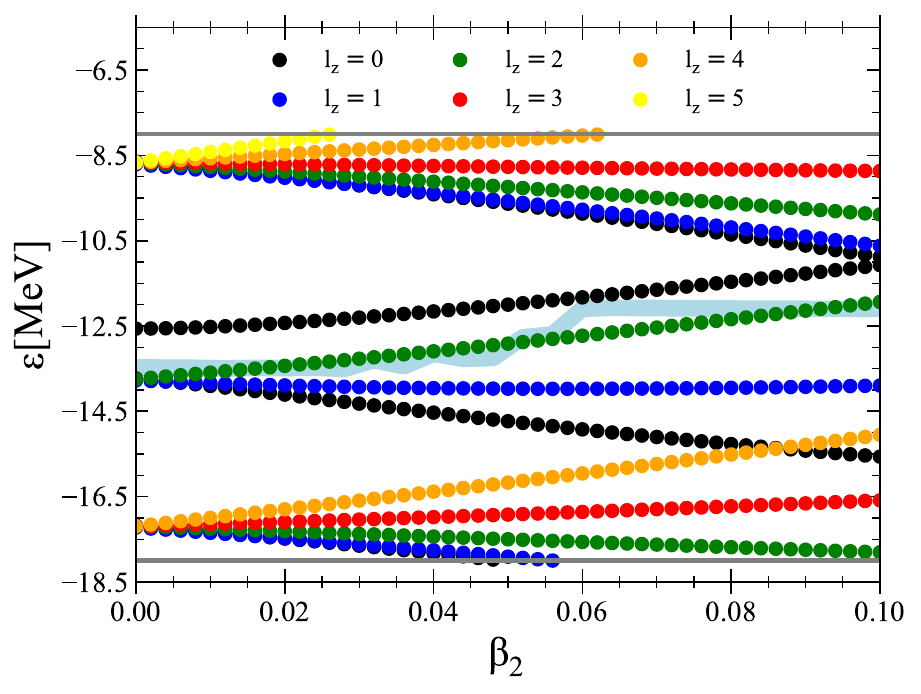}
\caption{The single particle states around the Fermi surface of $^{128}$Gd ($Z=64$). They are separated in $l_z$-shells which correspond, in ascending order, to $l_z=0$ (black), 1(blue), 2 (green), 3 (red), 4 (orange), and 5 (yellow). A thick blue line marks the chemical potential.}
\label{fig:lvl_gd}
\end{figure}

A closer look at the emergence of the mixed-spin pairing in $^{128}$Gd's ground state is given in Fig.~\ref{fig:128gd_constr} where, similarly to Fig.~\ref{fig:110cs_constr}, we plot the correlation energies of two constrained HFB states: one is kept to purely spin-singlet pairing (green squares), while the other to purely spin-triplet pairing (red diamonds). The correlation energy of the unconstrained state (i.e., the ground state) is also shown (blue solid line). The sudden amplification of the spin-orbit field at $\beta_2\approx0.06$ that is seen in Fig.~\ref{fig:128gd} depletes the spin-triplet correlation energy enough to bring it slightly lower than the spin-singlet correlation energy. The proximity of the two correlation energies is the sufficient condition for the emergence of the mixed-spin pairing correlations that dominate the ground state at this deformation and are marked by a dashed blue line in the plot. For $\beta_2>0.1$, the spin-singlet pairing correlation energy is further suppressed by deformation while the spin-triplet one is only slightly modulated by the oscillating spin-orbit field seen in Fig.~\ref{fig:128gd}, in the bottom panel, and spin-triplet pairing takes over. In Fig.~\ref{fig:lvl_gd} we plot the single-particle states around  $^{128}$Gd's Fermi surface which point to the underlying mechanism for the aforementioned increase in the spin-orbit field's strength.  With increasing $\beta_2$ deformation, many $l=5$ states, are shifted to lower energies approaching the Fermi surface marked by a light blue solid line in Fig.~\ref{fig:lvl_gd}. These states lie closer to the nuclear surface, where the spin-orbit field peaks, and as they get populated they cause the strengthening of the spin-orbit field seen in Fig.~\ref{fig:128gd}.

Going back once again to the systematics of Fig.~\ref{fig:nz0_b2all}, turning to higher deformation, at $\beta_2=0.25$, all nuclei in $N=Z$  line in the region, that have robust pairing correlations, exhibit spin-triplet pairing in their ground states. The peak in correlation energies seen at the spherical limit is now washed-out yielding a relatively uniform distribution of correlation energies. The evolution seen in Fig.~\ref{fig:nz0_b2all} points to a conclusion: quadrupole deformation's net effect in the region is to uniformly suppress pairing correlations but it favors spin-triplet pairing compared to spin-singlet through the mechanisms seen in Figs.~\ref{fig:110cs_constr} and \ref{fig:128gd_constr}.


\paragraph{Finite isospin asymmetry $N-Z$}
Moving away from the $N=Z$ line, in slices of fixed isospin asymmetries, we show correlation energies for $N-Z=1,2,5$ in Fig. ~\ref{fig:nz125_b2all}. Starting with the smaller isospin asymmetries, in Fig. \ref{fig:nz125_b2all}, filled, half-full, and empty points correspond, in the same order, to $N-Z=1,2,5$, and we keep the same colour- and shape-coding for the spin-symmetries as in Fig.~\ref{fig:nz0_b2all}. The dominant, and typical, effect of the isospin asymmetry is to raise the Fermi surface of the neutrons making proton-neutron pairing correlations harder to form. A lesser-recognized effect is that with well-separated Fermi surfaces, protons and neutrons move in different orbits with different degrees of proximity to the nuclear surface. As a result they experience different deformation effects. For instance, at the spherical limit, at $N-Z=2$ the difference between the neutron and proton chemical potentials for the nuclei with dominant spin-triplet pairing correlations is about $1.5$ MeV, but proton-neutron pairing correlations can still form. However, once the chemical potential difference reaches about $2$ MeV at $N-Z=12$, spin-triplet pairing correlations are no longer energetically favorable across such widely separated Fermi surfaces and spin-singlet pairing correlations dominate as those form between particles of the same species. In between these two extreme cases, the difference in Fermi surfaces weakens spin-triplet pairing correlations enough to allow mixed-spin ones (see open symbols in Fig.~\ref{fig:nz125_b2all}.

Once deformation is turned on, the isospin asymmetry retains the same effect albeit amplified and a small separation in Fermi surfaces is enough to weaken spin-triplet pairing correlations substantially. For small deformations, e.g., $\beta_2=0.1$, already at $N-Z=2$ or neutron and proton chemical potential difference of about $0.5$ MeV, spin-triplet pairing correlations are quenched giving way to dominant spin-singlet ones. We see the same picture at larger $\beta_2$ deformations. It is worth noting that at larger isospin asymmetries, at both deformations $\beta_2=0.1$ and $\beta_2=0.25$, the difference between neutron and proton chemical potentials is higher than the spherical limit. This can be attributed to the aforementioned difference of the deformation's effect on the pairing within the two nuclear species. In more detail, the well-separated Fermi surfaces mean that substantially different sets of single-particle states for each nuclear species partake in the pairing: the protons, whose Fermi surface remains close to what it was at the $N=Z$ line still pair in states of low-$l$ staying away from the Fermi surface while neutrons that must now pair in higher-$l$ states get closer to the nuclear surface where they are affected more by the surface's deformation. As a result the neutron-neutron pairing correlations are suppressed more than the proton-proton ones which makes their chemical potentials separate more than in the absence of deformation.

This detailed study of the response of the spin-singlet, spin-triplet, and mixed-spin pairing correlations to the quadrupole deformation reveals the main effects at play. While deformation suppresses pairing correlations, it has unequal effects on the different spin-symmetries favoring spin-triplet correlations most of the time. This is because the spin-triplet pairing correlations when dominant are less affected by the deformation as they are flushed to the interior of the nucleus by the spin-orbit field. Due to this unequal treatment, spin-singlet pairing suffers more from the deformation's averse effects and spin-triplet correlations often take the lead, especially on the $N=Z$ line where the Fermi surfaces of the two nuclear species are the closest and pairing across them the easiest. These effects do not depend strongly on the type of deformation and so even though they were identified by varying $\beta_2$, they can be anticipated for higher deformation modes as well.

\subsubsection{Higher deformation modes and odd-even mass staggering}
We now turn to higher deformation modes ($\beta_\lambda$ with $\lambda>2$) whose interplay with spin-triplet and mixed-spin pairing correlations has been hardly touched before in the literature. We will focus on a single nucleus, Eu ($Z=63$), and its isotopic chain for $A=126-136$. The quadrupole deformation prescribed by M{\"o}ller \textit{et al} in Ref.~\cite{ref:moller} changes only slightly for these nuclides, starting at $\beta_2=0.34$ for $\prescript{126}{63}{\textrm{Eu}}$ on the $N=Z$ line and decreasing steadily to $\beta_2=0.32$ for $\prescript{136}{63}{\textrm{Eu}}$. Regarding higher deformation modes, Ref.~\cite{ref:moller} predicts that these isotopes display no $\beta_3$ deformation, a $\beta_4$ deformation that is linearly decreasing with the neutron number, and $\beta_6$ deformation with a similar trend. Hence, by looking at realistically deformed Eu isotopes, we can see the effect of the higher deformation modes on top of an almost constant quadrupole deformation.

\begin{figure}
\includegraphics[width=0.5\textwidth]{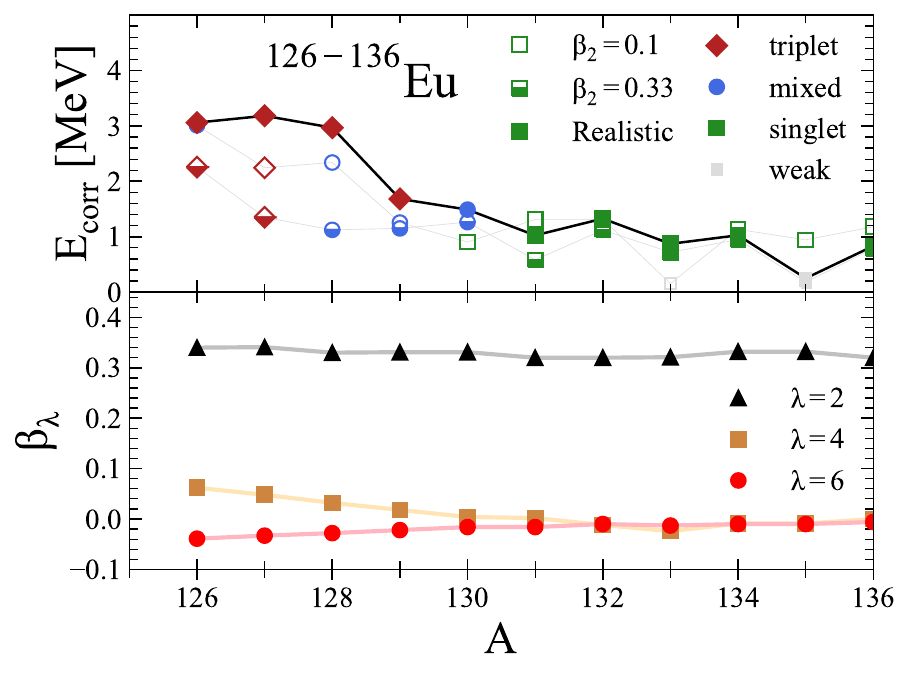}
\caption{Top panel: The correlation energies and symmetry of pairing correlations of the Eu (Z=63) isotopes for three different deformations: $\beta_2=0.1,\beta_2=0.33$, and the realistic deformation prescribed by Ref.~\cite{ref:moller}, marked by open, half-full, and full symbols, respectively. The correlation energies of the realistically deformed isotopes are additionally marked by a solid black line to guide the eye and signify its relevance to the deformation parameters shown on the bottom panel. Bottom panel: the deformation parameters of the Eu isotopes as prescribed by Ref.~\cite{ref:moller}.}
\label{fig:eu_moller}
\end{figure}

\begin{figure}
\includegraphics[width=0.5\textwidth]{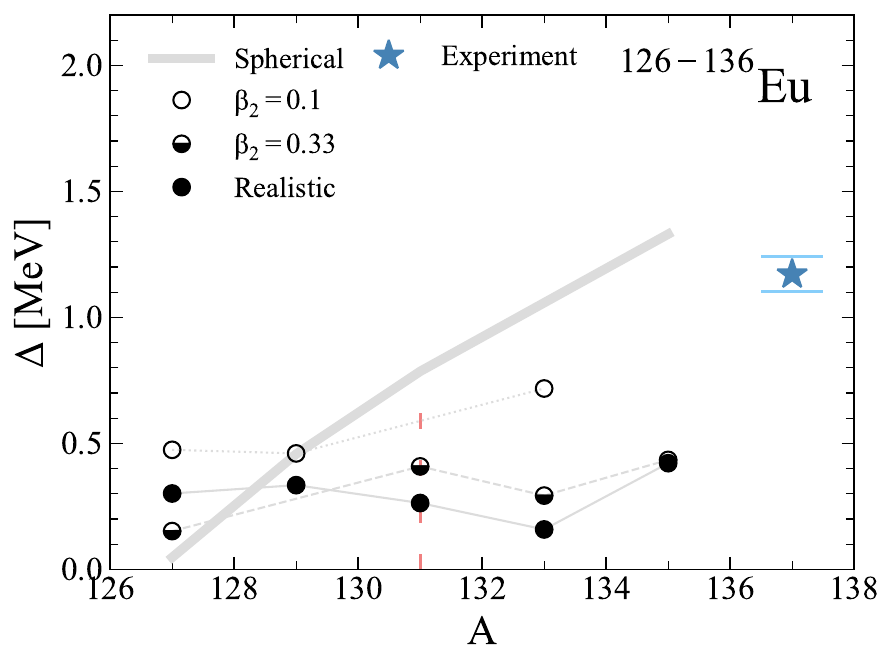}
\caption{Proton pairing gaps via the odd-even mass staggering for the Europium isotopes. The thick gray solid line corresponds to the spherical limit and the open, half-full, and full points correspond to small quadrupole, moderate quadrupole, and realistic deformation, respectively, with lines to guide the eye. The vertical red dashed line marks the onset of spin-singlet pairing as encountered  starting from the $N=Z$ line with realistic deformation (see Fig.~\ref{fig:eu_moller}).}
\label{fig:eu_pgaps}
\end{figure}

We studied quadrupole deformation for individual nuclei as well as regions of the nuclear chart in detail in Sec.~\ref{sec:beta_2} and disentangled the effects of the mass and the isospin asymmetry from that of deformation by changing one parameter at a time. This allows us to now turn to realistic deformation changing both the deformation parameters $\beta_4$, and $\beta_6$, and the neutron number at the same time, and look at their effects on the correlation energy. In Fig.~\ref{fig:eu_moller} the top panel shows the pairing correlations in the ground states of the isotopic chain of Eu with the y-axis measuring the correlation energies and the color- and shape-coding demonstrating the pairing's spin-symmetry. We show calculations with constant $\beta_2=0.1$ (open symbols) and $0.33$ (half-full symbols), the avergae quadrupole deformation for this part Eu's isotopic chain, and finally with the realistic deformation (full symbols with solid balck line) predicted by Ref.~\cite{ref:moller} which includes finite $\beta_2$, $\beta_4$, and $\beta_6$ deformations. The bottom panel refers to the latter case and shows the dependence of the deformation parameters on the mass of the isotopes in the chain.

The correlation energy of Eu at the spherical limit, which is not shown in Fig.~\ref{fig:eu_moller}, is $\sim10$MeV for the isotopes closest to the $N=Z$ line and it drops to $\sim2$ MeV for mass $A=134$ and higher. Looking at the open and half-full symbols in Fig.~\ref{fig:eu_moller} this correlation energy drops substantially as $\beta_2$ deformation increases; the mechanisms underlying this we discussed in sec.~\ref{sec:beta_2}. Also discussed in sec.~\ref{sec:beta_2} is how spin-triplet pairing correlations survive the deformation which is seen in Fig.~\ref{fig:eu_moller} too. A striking new feature is that higher deformation modes increase the correlation energy telling of an enhancement in the pairing correlations. This is demonstrated by the correlation energies for realistic deformation (full symbols in top panel) of Fig.~\ref{fig:eu_moller} which are consistently higher than the simply quadrupole-deformed ones with $\beta_2=0.33$, the average quadrupole deformation of these isotopes. For $A\le 130$ the enhancement is larger and it manages to counter a sizeable part of the suppression induced by the quadrupole deformation. This adds nuance to the known detrimental effect of deformation on pairing correlations: not all harmonic components of a nucleus' deformation [in the language of Eq.~(\ref{eq:cassini})] reduce pairing correlation.

The realistic deformation explored in this section brings our phenomenology the closest to what one would expect to see in the lab. In the same spirit, we now turn to the odd-even mass staggering, the traditional smoking gun of pairing correlations in mass measurements. In Fig.~\ref{fig:eu_pgaps} we show the proton pairing gaps calculated from Eq.~(\ref{eq:oes}) for the Europium isotopes. The open, half-full, and full symbols correspond to small ($\beta_2=0.1$), moderate ($\beta_2=0.33$), and realistic deformation, while the spherical limit is included as a solid line. Mass measurements in the region of light Lanthanides are scarce because these isotopes tend to be short lived with few-second half-lives. The current experimental knowledge stops some mass units away from where spin-triplet pairing correlations are expected to form and it tells of an average (spin-singlet) pairing gap about $\sim1.2$ MeV. We take this measurement as indicative of the magnitude of the spin-singlet pairing gaps in this region and mark it in Fig.~\ref{fig:eu_pgaps} by a blue star including its associated error bar.

The suppression of pairing gaps in the presence of spin-triplet proton-neutron pairing correlations has been proposed as a detectable effect of this elusive type of pairing~\cite{bertsch:2011}. Even though it has been explored multiple times in the past~\cite{gezerlis:2011,bulthuis:2016}, it has always been at the spherical limit raising questions about its validity as a probe. However, as shown in Fig.~\ref{fig:eu_pgaps}, we find that at any deformation, nuclei with spin-triplet correlations have reduced pairing gaps compared to the experimental value outside the region. Moreover, once the spin-singlet correlations take-over, for $A>131$, marked by the red dashed line, the pairing gaps trend upwards hinting that the observed suppression is not caused by the deformation alone. Nuclei with isospin aymmetries larger than the ones in Fig.~\ref{fig:eu_pgaps}, that is $N-Z>10$, have neutron chemical potentials that approach the high end of the regulating window ($-8~\textrm{MeV}$) described in sec.~\ref{sec:tune} laying most of the pair occupations on states outside the window. To extend our description to such systems requires widening the regulating window, refitting the effective contact interaction, and solving the HFB equations in a much larger single-particle space which is computationally challenging. A generalization of our HFB formulation that goes beyond an effective contact interaction while considering the different pairing channels studied here is beyond the scope of this paper and its left for future work.


\section{Discussion and Conclusions}

Both pairing correlations and deformation are emergent properties of finite nuclear systems and the relation between the two is long-standing. We have studied their interplay by separating them: we probed the response of different types of pairing correlations to different types of deformation. This is especially relevant to the region of the lightest Lanthanides where pairing correlations of various spin-symmetries are expected to form alongside axial deformation of different multipoles.

We have presented a formulation of the traditional HFB method that can handle any type of axial deformation using Cassini ovals and pairing correlations with multiple types of spin-symmetry. We applied the resulting deformed multimodal HFB theory in calculating correlation energies and pairing gaps for the spin-singlet, spin-triplet, and mixed-spin pairing correlations. We studied carefully the effect of different types of deformation modes on the spin-symmetry of the pairing correlations. This is the machinery that lead to the results of Ref.~\cite{Palkanoglou:2025} where it was found that deformation enhances spin-triplet pairing in the ground states of the lightest Lanthanides, close to $N=Z$. 

For simply quadrupole deformation, we found that the $\beta_2$ value expected in the region tips the scale in the competition between spin-singlet and spin-triplet correlations towards the latter. The underlying mechanism relies on the single-particle level structure at the Fermi surface, where pairing correlations peak,  and how that responds to deformation. The spin-orbit field, which resides on the nuclear surface, ensures that the spin-triplet pairing correlations form mainly in the interior of the nucleus, being formed between particles in low-$l$ orbitals. Hence, small deformation affects less this type of pairing unless it brings higher-$l$ single particle states close to the Fermi surface. The latter manifests as an increase in the spin-orbit field strength originating from single-particle wavefunctions that are sizeable at the nuclear surface. By studying the evolution of the single-particle energies with the increase of the quadrupole deformation, we find that states with low-$l$ quantum numbers at the spherical limit split in $2l+1$ states most of which move downwards (in the energy's absolute value) with increasing $\beta_2$ and this is also the trend of the chemical potential. The latter results from the deformation's general weakening of the pairing correlations induced by the decreased degeneracy of the single-particle states. Assuming that the low-$l$ single particle states retain the property of having spatial wavefunctions far from the nuclear surface even when deformed (this is reasonable for small to moderate deformation, like the one predicted for the Lanthanides' region), having those close to the Fermi surface ensures that the pairs formed on them lie close to the center of the nucleus. These pairs are less affected by the spin-orbit field and they can form spin-triplet pairing.

For deformations with higher modes, the picture in the Lanthanides' region remains qualitatively similar. This is because the higher-multipole deformation predicted for the region is relatively small, and the moderate $\beta_2$ deformation dominates. A new effect is that the higher-multipoles seem to enhance the pairing correlations which appends to the familiar suppression that deformation has on pairing correlations: not all harmonic components of deformation are detrimental to pairing correlations. For these more involved deformations we also calculated proton pairing gaps via the odd-even mass staggering which is experimentally accessible. The magnitude of the gaps appears reduced in the presence of spin-triplet pairing correlations when compared with the expected magnitude of spin-singlet pairing gaps in the region, solidifying this suppression as a fingerprint of spin-triplet pairing.

A few conclusions can be distilled from these results. First, as already reported in Ref.~\cite{Palkanoglou:2025}, spin-triplet pairing correlations are not destroyed by deformation but rather assisted in their competition with spin-singlet ones. This we interpret in detail here as ultimately originating from the spatial configuration of these pairing correlations. Second, the suppressed pairing gaps, a possible signature of the spin-triplet pairing correlations is seen as distinct from the suppression induced by deformation and hence still remains valid for realistic nuclei. Other signatures include enhanced neutron-proton transfer reaction cross-sections and similarities in spectroscopic properties of even-even and odd-odd nuclei~\cite{Frauendorf:2001}. While the the former might be prohibitively hard for this region of the nuclear chart, the latter remains relatively unexplored especially for the case of mixed-spin pairing and we leave it as a potential next step.

Throughout this study we have considered the nuclear deformation as fixed and only investigated the pairing correlations it creates. As mentioned before, we do so because it provides insight: deformation and pairing are emergent properties of nuclei and in principle should be considered on equal footing. This means that, when all nuclear effects are accounted for, a nucleus in its ground state will develop the deformation and pairing correlations that minimize its energy. Insofar as this problem cannot be tackled yet, one can at most separate the two and study their interplay. Therefore, one could investigate the type of deformation that certain pairing correlations create, i.e., the response of deformation to the pairing. From our results then, we would anticipate that spin-triplet pairing could favor moderate to high nuclear deformation. We can then interpret deformation of that magnitude, when it appears in heavy nuclei close to the $N=Z$ line, as a hint for underlying spin-triplet pairing correlations.

The structure of non-spin-singlet pairing correlations seen in finite nuclear systems remains rich and largely unexplored. This work opens a new window in their properties looking at their interface with deformation. At the same time, the deformed multimodal HFB developed here can produce mean-field states with more static correlations than deformed HFB states without neutron-proton pairing correlations, or spherical neutron-proton HFB states, which are often used to capture the static correlations associated with breaking a $U(1)$ symmetry in various \textit{ab initio} approaches~\cite{Scalesi:2024,Tichai:2018,Hergert:2018}. In nuclei such as the lightest Lanthanides where spin-triplet and mixed-spin pairing correlations are important, capturing them in a static way might lead to a robust first-principles description. Finally, the role of pairing correlations in more exotic heavy and deformed finite systems, like the ones arising in scission, remains to be seen. 

\

\section{Acknowledgments}
We thank  P.~Garrett, E.~Leistenschneider, and T.~Papenbrock for useful discussions. We also thank K. Neerg{\aa}rd for his careful reading of the manuscript. This work was supported by the Natural Sciences and Engineering Research Council (NSERC) of Canada and the Canada Foundation for Innovation (CFI). TRIUMF receives federal funding via a contribution agreement with the National Research Council of Canada. Computational resources were provided by SHARCNET and NERSC.

\end{document}